\documentclass{ametsocV6.1_no_lineno}

\title{Statistical Response of ENSO Complexity to Initial Condition and Model Parameter Perturbations}

\authors{Marios Andreou\aff{a}\correspondingauthor{Marios Andreou, mandreou@math.wisc.edu} and Nan Chen\aff{a}}

\affiliation{\aff{a}{Department of Mathematics, University of Wisconsin-Madison, Madison, WI, USA}}

\abstract{Studying the response of a climate system to perturbations has practical significance. Standard methods in computing the trajectory-wise deviation caused by perturbations may suffer from the chaotic nature that makes the model error dominate the true response after a short lead time. Statistical response, which computes the return described by the statistics, provides a systematic way of reaching robust outcomes with an appropriate quantification of the uncertainty and extreme events. In this paper, information theory is applied to compute the statistical response and find the most sensitive perturbation direction of different El Ni\~no-Southern Oscillation (ENSO) events to initial value and model parameter perturbations. Depending on the initial phase and the time horizon, different state variables contribute to the most sensitive perturbation direction. While initial perturbations in sea surface temperature (SST) and thermocline depth usually lead to the most significant response of SST at short- and long-range, respectively, initial adjustment of the zonal advection can be crucial to trigger strong statistical responses at medium-range around 5 to 7 months, especially at the transient phases between El Ni\~no and La Ni\~na. It is also shown that the response in the variance triggered by external random forcing perturbations, such as the wind bursts, often dominates the mean response, making the resulting most sensitive direction very different from the trajectory-wise methods. Finally, despite the strong non-Gaussian climatology distributions, using Gaussian approximations in the information theory is efficient and accurate for computing the statistical response, allowing the method to be applied to sophisticated operational systems.}

\begin{document}

\maketitle

%
%
%
\statement
The purpose of this work is to better understand how the El Niño-Southern Oscillation (ENSO) responds to changes in its initial state and internal dynamics or external forcings. A statistical quantification of this response allows for the comprehension of the triggering conditions and the effect of climate change in the occurrence frequency and strength of each type of ENSO event. Such a study also allows to detect the most sensitive perturbation directions, which has practical significance in guiding anthropogenic activities. The approach used to study the response in this work is through the framework of information theory, which allows for an unbiased and robust assessment of the statistical response that is not affected by the turbulent dynamics of the system.
%
%
%

%

\section{Introduction}
El Ni\~no-Southern Oscillation (ENSO) is the dominant interannual variability over the equatorial central to eastern Pacific Ocean \citep{philander1983nino, ropelewski1987global, klein1999remote, mcphaden2006enso}, which is characterized by its irregular and quasi-periodic anomalies in atmospheric wind and sea surface temperatures (SST). While the immediate effects are observed in the tropics and subtropics, ENSO has a significant impact on the climate, ecosystems, economies, and societies around the globe via atmospheric pathways \citep{ropelewski1987global, klein1999remote}, making it a global climate phenomenon leading to tropical cyclones, floods, and droughts. Under the classical viewpoint, ENSO is regarded as a climatological phenomenon with oscillatory behavior between two roughly mirror phases based on its features during its mature phase in the eastern Pacific (EP) \citep{jin1997equatorial}. In the warming phase of the SST, it is known as El Ni\~no, while in the cooling phase, it is known as La Ni\~na. In recent decades, many El Ni\~no events have been observed to occur in the central Pacific (CP) area. Therefore, the El Ni\~no phenomenon is further divided into the EP and CP types \citep{ashok2007nino, yu2007decadal, kao2009contrasting}, where the most significant SST anomaly (SSTa) is located near the coast of South America and the dateline, respectively \citep{yu2007decadal, kao2009contrasting}. This is known as the ENSO diversity \citep{capotondi2015understanding}. Although ENSO was initially viewed as an essentially symmetric phenomenon with oscillatory and quasi-periodic behavior, it demonstrates significant asymmetric spatial patterns and irregularities, as well as diverse characteristics in its temporal evolution and peak intensities. The former usually leads to different ENSO categories, including EP El Ni\~no, CP El Ni\~no, mixed EP-CP events, and La Ni\~na. The latter consists of standard single-year moderate ENSO events, multi-year events \citep{yu2018distinct}, extreme El Ni\~nos \citep{chen2015strong, levine2016extreme, capotondi2018nature, sun200910} and delayed El Ni\~nos \citep{hu2016exceptionally, hu2017extreme}. These spatiotemporal irregularities are called ENSO complexity \citep{timmermann2018nino, hayashi2017enso, boucharel2021influence}.

Due to the strong connections with other climate variabilities, it is of practical importance to compute the response of each ENSO event to various perturbations of the climate system. On the one hand, predicting the corresponding spatiotemporal patterns of an ENSO event due to the perturbed initial state helps understand its precursors and analyze the triggering conditions of the event. For example, calculating the response of the SST field due to the perturbed wind stress field is one of the most essential practical topics. Such a study is vital in advancing an improved understanding of ENSO predictability. It also facilitates discovering the formation mechanisms of extreme El Ni\~nos and thus provides possible guidance to prevent or cope with the associated natural hazards. On the other hand, a perturbation of either the internal dynamics or the external forcing will also lead to a change in the resulting ENSO spatiotemporal patterns. Such a perturbation can be caused by potential climate change and may result in regime switching. The perturbed system may bring about a different occurrence frequency for each type of ENSO event. It may also increase the strength of extreme events and enhance the probability of generating multi-year events. With an appropriate climate model, a natural way to study these two types of perturbation problems is to compute the model response to the perturbed initial conditions and model parameters, respectively. Particularly, detecting the most sensitive perturbation directions, namely the fastest-growing perturbations, has practical significance in guiding anthropogenic activities.

A hierarchy of approaches has been developed to study the response to these perturbations. One of the simplest methods for studying the initial perturbation is to approximate the nonlinear governing equation by the tangent linear model and then apply a linear singular vector to find the fastest-growing perturbation \citep{lorenz1965study, samelson2001instability}. To take into account crucial nonlinear features in computing the model response, methods based on the nonlinear singular vector have been built \citep{mu2000nonlinear, mu2001nonlinear}. The approaches were later extended to the development of conditional nonlinear optimal perturbation for studying the response of the ENSO and other climate phenomena \citep{mu2003new, duan2018predictability}. In addition, many studies exploited intermediate or global circulation models to carry out numerical experiments that compare the model realizations under various perturbations \citep{cai2020enso, mayer2016enso, toniazzo2008variation, callahan2021robust}. Most existing methods aim to study the trajectory-wise difference caused by perturbations. These methods provide valuable insights for short-term behavior and lead to many successful results. However, since nature is turbulent, applying trajectory-wise methods for studying the response may not always be an optimal approach. One of the fundamental issues in many complex turbulent systems is model error, which is inevitable in practice. Due to the turbulent nature, model error can easily be amplified after a very short term \citep{chen2023stochastic}. Consequently, model error dominates the intrinsic dynamics in driving the time evolution of the model trajectory. In other words, a large portion of the computed response is attributed to the model error. Furthermore, as stochastic parameterizations have been incorporated into many climate systems \citep{palmer2009stochastic, berner2017stochastic, franzke2015stochastic}, each single model trajectory contains randomness, which raises difficulties in using standard trajectory-wise approaches to evaluate the response.

Statistical response, which computes the deviation of the model statistics instead of trajectories, provides an alternative way to study the model response due to the perturbation of the initial state or parameters \citep{majda2018strategies, majda2019linear}. One significant advantage of the statistical response is that although each model trajectory is chaotic, the time evolution of the statistics is deterministic and more predictable \citep{gardiner1985handbook, chen2023stochastic}. These statistics are also robust to the random noises in the underlying system. As a result, a small model error will not substantially impact the response of the statistics, which is fundamentally different from its trajectory-wise counterpart. Notably, the statistical response not only describes the shift of the mean state but also measures the increase or decrease of the level of uncertainty. The latter can be roughly reflected in the variance or, more precisely, characterized by the probability density function (PDF) of the state variables \citep{billingsley2017probability}. Understanding the corresponding change of the uncertainty as a response to perturbations is essential in studying the predictability in ensemble prediction and advancing the probabilistic forecast of extreme events \citep{fang2023quantifying}. Although several statistical methods have been utilized to assess prediction skill, intrinsic predictability, and model error \citep{delsole2004predictability, delsole2007predictability, kleeman2011information, majda2005information}, analyzing the statistical response of ENSO complexity has not been systematically studied.

In this paper, a mathematical framework for computing the statistical response of a complex system is developed, where information theory is utilized to measure the strength of the response. The method is then applied to study the response of different ENSO events, namely the ENSO complexity, to the perturbations of initial conditions and model parameters. It is also used to find the most sensitive perturbation direction for each ENSO event. The focus is on highlighting the advantage of the statistical response over the trajectory-wise approaches, especially for computing the response in characterizing the uncertainty and extreme events. A recently developed conceptual multiscale stochastic model is utilized to describe the ENSO complexity \citep{chen2022multiscale}. The model can reproduce many crucial observed dynamical and statistical features, including the non-Gaussian climatology statistics. It has also been the building block for developing an intermediate coupled stochastic dynamical model for the ENSO complexity \citep{chen2023simple}. Therefore, the conceptual model is an appropriate choice as a first path for exploring the statistical response of different ENSO events. In addition to the direct numerical algorithm based on the definition, several approximate schemes are derived to facilitate the practical calculations of the statistical response. These methods are applicable to more sophisticated operational models. The conceptual model will serve as a testbed to validate these computationally efficient methods.

The rest of the paper is organized as follows. The general framework of the statistical response and the associated computationally efficient approximate schemes are introduced in Section \ref{Sec:Methods}. The observational data set and the conceptual multiscale stochastic model for the ENSO complexity are described in Section \ref{Sec:Obs_Model}. The results of the statistical response to initial value and model parameters are presented in Section \ref{Sec:Results}. The paper is concluded in Section \ref{Sec:Conclusion}.

\section{Methods}\label{Sec:Methods}
\subsection{Standard trajectory-wise response}
Let us start reviewing the standard trajectory-wise methods for computing the model response. Denote by $\mathbf{x}_t$ the state variable $\mathbf{x}$ at time $t$ from the original model. Denote by $\mathbf{x}^{\delta}_t$ the corresponding variable once a perturbation $\boldsymbol{\delta}$ is imposed on either the initial conditions or the model parameters. Note that, although the superscript $\delta$ in $\mathbf{x}^{\delta}_t$ is written in the scalar form for notational simplicity, the perturbation $\boldsymbol{\delta}$ is usually a vector. The response of the system at time $t$ to such a perturbation can be defined as the distance between these two vectors \citep{samelson2001instability}, namely
\begin{equation}\label{distance_pathwise}
d(\mathbf{x}^\delta_t,\mathbf{x}_t)=\|\mathbf{x}_t-\mathbf{x}^\delta_t\|,
\end{equation}
where a standard Euclidean norm is taken to obtain a scalar value of the distance.

Due to the chaotic nature of the system, one fundamental issue is that model error can dominate the intrinsic dynamics after a short term. Therefore, it remains unclear in many situations if the computed distance is attributed to the actual response or the model error.

\subsection{Statistical response and the associated information measurement}
Different from using the trajectories as the quantity for measuring the model response, the statistical response focuses on the change in the model statistics as a response to the perturbations. To this end, denote by $p(\mathbf{x}_t)$ the PDF of $\mathbf{x}_t$ for the original unperturbed system and $p^\delta(\mathbf{x}_t)$ the corresponding PDF after the perturbation. The moments, such as the mean and the variance, can be easily obtained from the PDF. Unlike the case with two vectors where the point-wise difference as in \eqref{distance_pathwise} can be naturally used as the distance, such a direct discrepancy between the two PDFs will significantly underestimate the role of extreme events corresponding to the tail probability. Information theory provides a practical way to characterize the distance between the two PDFs via the following relative entropy \citep{majda2010quantifying, majda2005information, kleeman2011information},
\begin{equation}\label{relative_entropy}
  \mathcal{P}(p^\delta(\mathbf{x}_t), p(\mathbf{x}_t))=\int_{\mathbf{x}_t} p^\delta(\mathbf{x}_t)\log\left(\frac{p^\delta(\mathbf{x}_t)}{p(\mathbf{x}_t)}\right)\mathrm{d} \mathbf{x}_t,
\end{equation}
which is also known as Kullback-Leibler divergence or information divergence \citep{kullback1951information, kullback1987letter, kullback1959statistics}. The ratio between the two PDFs inside the logarithm function quantifies the gap in the tail probability, resulting in an unbiased way of characterizing the statistical difference. It allows the relative entropy to be widely utilized to quantify model error, predictability, and prediction skill \citep{majda2010quantifying, majda2011link, majda2012lessons, branicki2012quantifying, branicki2014quantifying, kleeman2011information, kleeman2002measuring, delsole2004predictability, branicki2013non, branstator2010two}.
Despite the lack of symmetry, the relative entropy has two attractive features. First, $\mathcal{P}(p^\delta(\mathbf{x}_t), p(\mathbf{x}_t)) \geq 0$ with equality if and only if $p^\delta(\mathbf{x}_t)= p(\mathbf{x}_t)$. Second, $\mathcal{P}(p^\delta(\mathbf{x}_t), p(\mathbf{x}_t))$ is invariant under general nonlinear changes of variables. These provide an attractive framework for assessing the discrepancy between the two statistical quantities. A larger value of $\mathcal{P}(p^\delta(\mathbf{x}_t), p(\mathbf{x}_t))$ means the statistical response to the perturbation is more significant.

Since the relative entropy $\mathcal{P}$ in \eqref{relative_entropy} is unbounded, it is practically useful to introduce a rescaled version defined as
\begin{equation}\label{Rescaled_Relative_Entropy}
  \mathcal{E} = 1-\exp(-\mathcal{P}),
\end{equation}
It rescales the original relative entropy $\mathcal{P}$ to the interval $[0, 1)$. The rescaled relative entropy $\mathcal{E}$ takes the value of $0$ if and only if $\mathcal{P} = 0$. It approaches $1$ when $\mathcal{P}$ becomes infinity. The rescaled relative entropy $\mathcal{E}$ remains a monotonically increasing function in terms of the difference between the two PDFs. The rescaled relative entropy $\mathcal{E}$ will be used in all the numerical results shown in this work.

Given a perturbation, the strength of the corresponding statistical response is computed from the relative entropy in \eqref{relative_entropy}. However, unless the distributions have desirable features, numerical integration is needed to calculate the relative entropy, which is a computationally challenging issue. Furthermore, in the situation of seeking the most sensitive perturbation directions, an exhaustive search of the entire state space of $\mathbf{x}_t$ is needed based on the direct definition of the relative entropy in \eqref{relative_entropy}. This becomes computationally prohibitive when $\mathbf{x}_t$ is high dimensional \citep{robert2010introducing, kuo2005lifting}. The following subsection aims to provide alternative ways to accelerate the calculations.

\subsection{Practical numerical approaches for computing the statistical response} \label{Subsec:PracticalApproaches}
\subsubsection{Gaussian approximation}
One practical setup for utilizing the framework of information theory in many applications arises when both the distributions are Gaussian so that $p^\delta(\mathbf{x}_t)\sim\mathcal{N}(\bar{\mathbf{x}}_t^{\delta}, \mathbf{R}_t^{\delta})$ and $p(\mathbf{x}_t)\sim\mathcal{N}(\bar{\mathbf{x}}_t, \mathbf{R}_t)$. In the Gaussian framework, $\mathcal{P}(p^\delta(\mathbf{x}_t), p(\mathbf{x}_t))$ has the following explicit formula \citep{majda2010quantifying, majda2006nonlinear}
\begin{equation}\label{Signal_Dispersion}
\begin{split}
  \mathcal{P}(p^\delta(\mathbf{x}_t), p(\mathbf{x}_t)) =& \left[\frac{1}{2}(\bar{\mathbf{x}}^\delta_{t}-\bar{\mathbf{x}}_{t})^\mathtt{T}(\mathbf{R}_{t})^{-1}(\bar{\mathbf{x}}^\delta_{t}-\bar{\mathbf{x}}_{t})\right] \\&\qquad+ \left[-\frac{1}{2}\log\det(\mathbf{R}^\delta_{t}\mathbf{R}_{t}^{-1}) + \frac{1}{2}(\mbox{tr}(\mathbf{R}^\delta_{t}\mathbf{R}_{t}^{-1})-\mbox{Dim}(\mathbf{x}_t))\right],
\end{split}
\end{equation}
where $\mbox{Dim}(\mathbf{x}_t)$ is the dimension of $\mathbf{x}_t$.
The first term in brackets in \eqref{Signal_Dispersion} is called `signal', reflecting the information gain in the mean but weighted by the inverse of the model variance, $\mathbf{R}_{t}$, whereas the second term in brackets, called `dispersion', involves only the covariance ratio, $\mathbf{R}^\delta_{t}\mathbf{R}^{-1}_{t}$. The signal and dispersion terms are individually invariant under any (linear) change of variables which maps Gaussian distributions to Gaussians.

For non-Gaussian PDFs, a Gaussian fit using the mean and covariance can always be adopted to build the approximate Gaussian distributions. Then the explicit formula in \eqref{Signal_Dispersion} is used to find the approximate statistical response. It is worth highlighting two things. First, the Gaussian approximation in \eqref{Signal_Dispersion} is very different from using a linear approximation of the original dynamics, such as the linear tangent model. The full nonlinear model is still utilized to obtain the non-Gaussian PDF as the first step. Only the Gaussian statistics of the non-Gaussian distribution are used in the explicit formula \eqref{Signal_Dispersion}. Therefore, the statistical response still reflects the nonlinear features of the underlying dynamics. Second, although the Gaussian approximation may lead to errors in approximating the PDF itself, it may become a valuable surrogate for finding the most sensitive perturbation direction, corresponding to the strongest statistical response at a given forecast lead time. Therefore, one task below compares the statistical response computed from the explicit formula with a Gaussian approximation with the exact value. The conclusions based on the conceptual model tests can provide valuable guidelines for more sophisticated models.

\subsubsection{Leading-order approximation via Fisher information}
Recall that $\boldsymbol\delta$ is the perturbation vector, which can be the perturbation of a subset of the state variables for the initial values or a few selected parameters, and assumed to be an $N$ dimensional vector. It is assumed that the possible range for the perturbation is within physical meanings and that it is further standardized as for $\boldsymbol\delta=\mathbf{0}$ to correspond to the unperturbed climatological system. Since the perturbation is usually small, the perturbed PDF can be written as a function of $\boldsymbol\delta$. Applying a Taylor expansion of $p^\delta(\mathbf{x}_t)$ with respect to $\boldsymbol\delta$ in computing the relative entropy under the tacit assumption that the PDF is differentiable with respect to the perturbation $\boldsymbol \delta$ \citep{majda2009high, majda2010linear, hairer2010simple}, yields the following leading-order approximation of the response to the perturbation \citep{majda2018model, majda2011link},
\begin{equation}\label{quadratic_form}
  \mathcal{P}(p^\delta(\mathbf{x}_t), p(\mathbf{x}_t)) = \frac{1}{2}\boldsymbol\delta\cdot I(p(\mathbf{x}_t))\boldsymbol\delta + O(|\boldsymbol\delta|^3),
\end{equation}
where the first term on the right-hand side of \eqref{quadratic_form} is the quadratic form in $\boldsymbol\delta$ defined by the $N\times N$ Fisher information matrix \citep{williams2001weighing, cover1999elements},
\begin{equation}\label{Fisher_Info}
  \boldsymbol\delta\cdot I(p(\mathbf{x}_t))\boldsymbol\delta = \int_{\mathbf{x}_t}\frac{(\boldsymbol\delta\cdot\nabla_{\boldsymbol\delta}p(\mathbf{x}_t))^2}{p(\mathbf{x}_t)}\mathrm{d}\mathbf{x}_t,
\end{equation}
where the gradients are evaluated at the unperturbed state. 

One significant advantage of the quadratic form in \eqref{Fisher_Info} is that the most sensitive perturbation direction, namely the strongest statistical response at $t$, occurs along the unit direction associated with the largest eigenvalue of the matrix $I(p(\mathbf{x}_t))$. Such an eigenvalue can be easily computed once the gradients of $p(\mathbf{x}_t)$ along the directions of the basis vectors of $\boldsymbol\delta$ are calculated, which requires only a small number of evaluations. In contrast, the computationally expensive brute-force search algorithms, computing the response at all possible directions, have to apply when the exact formula in \eqref{relative_entropy} or its Gaussian approximation \eqref{Signal_Dispersion} is used to achieve such a goal. Therefore, the Fisher information provides an efficient and systematic way in finding the most sensitive perturbation.

\subsubsection{Fisher information with coarse-grained statistical measurements}
In many situations, the observed climatology data are used to compute unperturbed statistics. Yet, it is worth noting that, due to the limited amount of data and the possible measurement noise, these data can usually estimate the first few moments accurately, but the higher-order moments are very sensitive to the small noise. Therefore, instead of calculating the exact PDF, the measured leading few moments are typically used to reconstruct the least biased PDF using the so-called maximum entropy principle \citep{majda2006nonlinear, bajkova1992generalization}. It is then used as an approximation of $p(\mathbf{x}_t)$ in \eqref{Fisher_Info} to compute the most sensitive perturbation direction. See \cite{majda2005information}, \cite{majda2006nonlinear}, and \cite{majda2010quantifying} for more details. 

Denote by $\mathbf{E}_L(\mathbf{x}_t) = (E_1(\mathbf{x}_t),\ldots, E_L(\mathbf{x}_t))$ the $L$ statistical quantities from the observational measurements or model simulation, for example, the mean, the covariance and up to the $L$-th moment. The least biased PDF obtained from the maximum entropy principle is denoted by $p^{{\delta}}_L(\mathbf{x}_t)$. Further denote by $\overline{\mathbf{E}}_L = (\overline{E}_1,\ldots, \overline{E}_L)^\mathtt{\, T}$, where each component $\overline{{E}}_l$ is given by $\overline{{E}}_l = \int_{\mathbf{x}_t}{E}_l(\mathbf{x}_t)p^{{\delta}}_L(\mathbf{x}_t)\mathrm{d} \mathbf{x}_t$. The notation `overline' represents the statistical average with respect to the perturbed PDF, and the resulting $\overline{{E}}_l$ is a number that depends on the perturbed dynamics. Then the quadratic form of $\boldsymbol\delta \cdot I(p(\mathbf{x}_t))\boldsymbol\delta$ in \eqref{quadratic_form} can be approximated by \citep{majda2010quantifying}
\begin{equation}\label{quadratic_form_approx}
  \boldsymbol\delta\cdot I(p_L(\mathbf{x}_t))\boldsymbol\delta = \boldsymbol\delta\cdot\left(\big(\nabla_{\boldsymbol\delta}\overline{\mathbf{E}}_L\big)^\mathtt{\, T}\mathcal{C}^{-1}\nabla_{\boldsymbol\delta}\overline{\mathbf{E}}_L\right)\boldsymbol\delta,
\end{equation}
where $\mathcal{C}$ is the $L\times L$ climate correlation matrix
\begin{equation}\label{climate_correlation_matrx}
  \mathcal{C} = \overline{(\mathbf{E}_L(\mathbf{x}_t)-\overline{\mathbf{E}}_L)(\mathbf{E}_L(\mathbf{x}_t)-\overline{\mathbf{E}}_L)^\mathtt{T}},
\end{equation}
and $\nabla_\delta\overline{\mathbf{E}}_L$ is the gradient of each component in $\overline{\mathbf{E}}_L$ with respect to the perturbation vector $\boldsymbol{\delta}$ that gives an $L\times N$ Jacobi matrix with standard element $\displaystyle\left(\nabla_\delta\overline{\mathbf{E}}_L\right)^{j=1,\ldots,N}_{l=1,\ldots,L}=\frac{\partial \overline{{E}}_l}{\partial \delta_j}$ and with the Jacobian of the statistical average being evaluated at the unperturbed state of $\boldsymbol{\delta}=\mathbf{0}$. Note that ${E}_l(\mathbf{x}_t)$ is a function of the state variable $\mathbf{x}_t$ while its statistical average $\overline{{E}}_l$ is a number. For example, if the mean and the variance are adopted as the first two components of the measurements, then ${E}_1(\mathbf{x}_t)=\mathbf{x}_t$ and ${E}_2(\mathbf{x}_t)=(\mathbf{x}_t - \overline{\mathbf{x}}_t)^2$. Correspondingly, $\overline{E}_1(\mathbf{x}_t)=\overline{\mathbf{x}}_t$ and $\overline{E}_2(\mathbf{x}_t)=\overline{(\mathbf{x}_t - \overline{\mathbf{x}}_t)^2}$. The associated first four entries of $\mathcal{C}$ in \eqref{climate_correlation_matrx} are given by
\begin{equation*}
\mathcal{C}_{11}=\overline{(\mathbf{x}_t - \overline{\mathbf{x}}_t)^2},\qquad \mathcal{C}_{12}=\mathcal{C}_{21}=\overline{(\mathbf{x}_t - \overline{\mathbf{x}}_t)^3}, \qquad\mbox{and}\qquad \mathcal{C}_{22}=\overline{(\mathbf{x}_t - \overline{\mathbf{x}}_t)^4}-(\overline{(\mathbf{x}_t - \overline{\mathbf{x}}_t)^2})^2.
\end{equation*}

Finally, it is also practically useful to compute the compressed quadratic form involving fewer measurements, $L' \leq L$,
\begin{equation}\label{quadratic_form_approx2}
  \boldsymbol\delta\cdot I(p_L(\mathbf{x}_t))\boldsymbol\delta = \boldsymbol\delta\cdot\left(\big(\nabla_{\boldsymbol\delta}\overline{\mathbf{E}}_{L'}\big)^\mathtt{\, T}\mathcal{C}^{-1}\nabla_{\boldsymbol\delta}\overline{\mathbf{E}}_{L'}\right)\boldsymbol\delta,
\end{equation}
where $\mathbf{E}_{L'}(\mathbf{x}_t) = (E_1(\mathbf{x}_t),\ldots, E_{L'}(\mathbf{x}_t),0,\ldots,0))^\mathtt{\, T}$. That is, the first $L'$ entries of $\mathbf{E}_{L'}(\mathbf{x}_t)$ are the same as those in $\mathbf{E}_L(\mathbf{x}_t)$ but the remaining entries are zero. The compressed quadratic form in \eqref{quadratic_form_approx2} is relevant in determining the important practical information regarding whether, for example, changes in the mean climate statistics alone determine the most sensitive directions of climate change.

\subsection{Summary}\label{Subsec:Summary}
In the following, the exact formula of computing the statistical response in \eqref{relative_entropy} and its two approximations, namely the Gaussian approximation \eqref{Signal_Dispersion} and the approximation via the Fisher information with the quadratic form \eqref{quadratic_form}--\eqref{Fisher_Info} will be applied to compute the statistical response. Three different computational approaches will be adopted using the Fisher information with the quadratic form to find the most sensitive perturbation direction. They are the exact PDF as in \eqref{quadratic_form}--\eqref{Fisher_Info} (hereafter ``exact quadratic form''), the quadratic form with measuring only the mean and covariance \eqref{quadratic_form_approx}--\eqref{climate_correlation_matrx} (hereafter ``quad form w/ mean and variance'') and the compressed quadratic form involving only the mean \eqref{quadratic_form_approx2} (hereafter ``quad form w/ mean only'').  In Section \ref{Sec:Results}, the intercomparison between these five different computational methods will be carried out.

Note that, although the quad form w/ mean and variance utilizes the Gaussian statistics, it differs from applying the Gaussian approximation \eqref{Signal_Dispersion} to the exact formula. The quadratic form via the Fisher information \eqref{Fisher_Info} is already an approximation in computing the statistical response since the Taylor expansion of the relative entropy in \eqref{quadratic_form} is utilized. Building upon this, the first two moments are adopted to replace the full PDF as a second approximation. Nevertheless, it is worth highlighting that the methods exploiting the quadratic form facilitate determining the most sensitive perturbation direction by finding the unit eigenvector corresponding to the largest eigenvalue of the matrix $I(p(\mathbf{x}_t))$ or $I(p_L(\mathbf{x}_t))$, hereafter named the maximal eigenvector.

The statistical quantities in this work are computed from an ensemble simulation. It is based on a Monte Carlo simulation with $3000$ ensemble members. Such an ensemble size is large enough to reproduce the strong non-Gaussian climatology PDFs for the 6-dimensional conceptual model described in Section \ref{Sec:Obs_Model}. The computational cost of using $3000$ ensemble members remains low for such a conceptual model. Therefore, it provides an accurate reference solution of the statistical response using the exact formula. It further allows us to compare the result with those using the Gaussian approximations and the quad form w/ mean and variance. Note that a small ensemble size is usually sufficient to reach reasonably accurate results when applying the Gaussian approximation or the Gaussian statistics in the quadratic form. Therefore, these approximate methods can compute the statistical response using more sophisticated and higher-dimensional operational systems. The unperturbed initial value is always given by a Gaussian distribution, centered at the observational values and equipped with a tiny variance of $10^{-4}$ along each direction. Adding such a small uncertainty to the initial value facilitates the numerical calculation of the relative entropy within a very short lead time. Other advanced techniques, such as the fluctuation-dissipation theorem \citep{majda2005information, kubo1966fluctuation}, can be embedded into the above methods to compute the statistical response more efficiently. These methods are helpful for more complicated systems but are not adopted here since the direct Monte Carlo simulation with $3000$ ensemble members is sufficient for the 6-dimensional conceptual model. Finally, when calculating $\nabla_{\boldsymbol\delta}p(\mathbf{x}_t)$ in \eqref{Fisher_Info}, where $p(\mathbf{x}_t)$ is the unperturbed distribution, the derivative is approximated by a second-order accurate centered finite difference. This is achieved by adding a small numerical perturbation $\boldsymbol\epsilon$ with $\|\boldsymbol\epsilon\|\ll \|\boldsymbol\delta\|$ to the initial conditions or model parameters, and then computing the associated PDF at time $t$. Notably, the amplitude of the numerical perturbation $\boldsymbol\epsilon$ is required to be small to maintain the accuracy of computing the derivative and guarantee the dominant role of the actual perturbation $\boldsymbol\delta$ in computing the response. Likewise, we compute another PDF at time $t$ corresponding to the perturbation $-\boldsymbol\epsilon$. These two PDFs are then used to approximate the derivative numerically as
\begin{equation*}
    \frac{p^{\epsilon}(\mathbf{x}_t)-p^{-\epsilon}(\mathbf{x}_t)}{2\epsilon}.
\end{equation*}

The primary issues to be addressed in this work are the following:
\begin{enumerate}
  \item Study the statistical response of different ENSO events, namely the ENSO complexity, to the perturbations of initial conditions and model parameters. Find the most sensitive perturbation direction for each type of ENSO event.
  \item Since the quad form w/ mean only resembles the deterministic trajectory-wise response, its difference compared with other methods can be utilized to reveal the crucial role of the uncertainty in affecting the model response.
  \item Exploit the skill of the computationally efficient methods involving Gaussian statistics that apply to operational systems.
\end{enumerate}

Note that the focus of analyzing the responses to the perturbations of initial conditions and model parameters can be different. Since the initial value perturbation will have an immediate impact at short- and medium-range lead times, studying the difference in the response of each single ENSO event will be highlighted. In contrast, the short-term behavior of the system may not be affected by a sudden change in the parameter values. Therefore, parameter perturbation focuses more on long-term behavior that affects the climatology.

\section{Observational Data Sets, Definitions of Different Types of ENSO Events and the Multiscale Model}\label{Sec:Obs_Model}
\subsection{Observational data sets}
In most of the studies in this work, the monthly ocean temperature and current data are from the GODAS reanalysis dataset \citep{behringer2004evaluation}. The thermocline depth along the equatorial Pacific is approximated from the potential temperature as the depth of the $20^\circ$C isotherm contour. The analysis period is 36 years, from the start of 1982 until the end of 2017. Anomalies presented in this study are calculated by removing the monthly mean climatology of the whole period. In this work, the Ni\~no 4 ($T_C$) and Ni\~no 3 ($T_E$) indices are the averages of the SSTa over the CP ($160^\circ$E-$150^\circ$W, $5^\circ$S-$5^\circ$N) and EP ($150^\circ$W-$90^\circ$W, $5^\circ$S-$5^\circ$N) regions, respectively. The $h_W$ index is the mean thermocline depth anomaly over the western Pacific region ($120^\circ$E-$180^\circ$, $5^\circ$S-$5^\circ$N), while the $u$ index is the mean mixed-layer zonal current in the CP region. In the last subsection of the results, the ERSSTv5 data \citep{huang2017extended}, which has a longer period, is utilized to compute the number of each type of ENSO event. This longer period of observed SST data allows us to compute the number of ENSO events in Table 1 while minimizing the statistical biases. The 36-year observational period defined by the GODAS dataset may cause more errors in the statistics of interest.

The daily zonal wind data is measured at 850 hPa and is taken from the NCEP-NCAR reanalysis \citep{kalnay1996ncep}. It is used to describe the wind bursts in the intraseasonal scale. Removing the daily mean climatology, the anomalies are averaged over the WP region to create the wind burst index. Note that the wind lies in a faster time scale than all other state variables (daily than monthly). Although a single daily value of the $\tau$ index describing the wind anomalies has a minor effect on the SST variables, the accumulated wind over time will modulate the SST variations. Jumping up to the decadal time scale, Walker circulation strength index data are also included to illustrate the modulation of the decadal variation on the interannual ENSO characters. It is defined as the sea level pressure difference over the CP/EP ($160^\circ$W-$80^\circ$W, $5^\circ$S-$5^\circ$N) and the Indian Ocean/WP ($80^\circ$E-$160^\circ$E, $5^\circ$S-$5^\circ$N) \citep{kang2020walker}. The monthly zonal SST gradient between the WP and CP region highly correlates with this Walker circulation strength index (with a simultaneous Pearson correlation coefficient of around 0.85), suggesting the significance of the air-sea interactions over the equatorial Pacific. Since the latter is more directly related to the zonal advective feedback strength over the CP region, the decadal model state variable ($I$) mainly illustrates this quantity.

\subsection{Definition of different types of ENSO events}
The definitions of different ENSO events are based on the average SSTa during boreal winter (DJF). The CP region is defined as $160^\circ$E-$150^\circ$W, $5^\circ$S-$5^\circ$N, with the former indicating the longitude (with $180^\circ$ being the Prime Meridian) and the latter the latitude (with $0^\circ$ being the Equator), and the EP region as $150^\circ$W-$90^\circ$W, $5^\circ$S-$5^\circ$N. Using the definitions of \citep{kug2009two}, when the EP is warmer than the CP and the EP SSTa, $T_E$, is greater than $0.5^\circ$C, it is classified as an EP El Ni\~no. Based on the classification in \citep{wang2019historical}, an extreme EP El Ni\~no event corresponds to when the maximum of the EP SSTa from April to the following March is larger than $2.5^\circ$C. Accordingly, when the CP is warmer than the EP and the CP SSTa, $T_C$, is larger than $0.5^\circ$C, it is defined as a CP El Ni\~no. Finally, when either the $T_C$ or $T_E$ anomalies are cooler than $-0.5^\circ$C, it is defined as a La Ni\~na event. 

\subsection{The multiscale stochastic dynamical model for the ENSO complexity}\label{Subsec:Model}
The model used to study the statistical response is a recently developed stochastic conceptual model for the ENSO complexity \citep{chen2022multiscale}:
\begin{subequations}\label{conceptual_model}
\begin{align}
    \cfrac{\mathrm{d} u}{\mathrm{d} t}&=-ru-\delta_u\cfrac{T_C+T_E}{2}+\beta_u(I)\tau+\sigma_u\dot{W}_u, \label{eq:adv}\\
    \cfrac{\mathrm{d} h_W}{\mathrm{d} t}&=-rh_W-\delta_h\cfrac{T_C+T_E}{2}+\beta_h(I)\tau+\sigma_h\dot{W}_h, \label{eq:thermocline}\\
    \cfrac{\mathrm{d} T_C}{\mathrm{d} t}&=(r_C-c_1(t,T_C))T_C+\zeta_CT_E+\gamma_Ch_W+\sigma(I)u+C_u+\beta_C(I)\tau+\sigma_C\dot{W}_C,\label{eq:sstac}\\
    \cfrac{\mathrm{d} T_E}{\mathrm{d} t}&= (r_E-c_2(t))T_E-\zeta_ET_C+\gamma_Eh_W+\beta_E(I)\tau+\sigma_E\dot{W}_E,\label{eq:sstae}\\
    \cfrac{\mathrm{d} \tau}{\mathrm{d} t}&=-d_\tau\tau+\sigma_\tau(t,T_C)\dot{W}_\tau, \label{eq:wind}\\
    \cfrac{\mathrm{d} I}{\mathrm{d} t}&=-\lambda(I-m)+\sigma_I(I)\dot{W}_I. \label{eq:walker}
\end{align}
\end{subequations}
The dimensional units and the parameters in the coupled model are summarised in Table 2 in the Appendix.

The dynamical core of the model is a deterministic three-region interannual linear model with zonal advective feedback \citep{fang2018three}. It extends on the classical two-region recharge oscillator model \citep{jin1997equatorial} and implements the air-sea interactions over the entire WP, CP, and EP regions. It also incorporates the ocean content discharge and recharge process controlling the occurrence of El Ni\~no and La Ni\~na events via the thermal layer and the ocean zonal advection. In the model, $T_C$ and $T_E$ are the SSTa in the CP and EP regions, respectively, while $u$ is the mean zonal current anomaly in the CP region and $h_W$ is the mean thermocline depth anomaly in the WP region. In addition to these interannual variabilities, two processes, describing the intraseasonal wind bursts $\tau$ and the decadal variability in the background Walker circulation $I$, are further incorporated into the model. The intraseasonal variability $\tau$ accounts for several important atmospheric ENSO triggers, such as the westerly wind bursts (WWBs), the easterly wind bursts (EWBs), and the convective envelope of the Madden-Julian Oscillation (MJO) \citep{chen2015strong, hu2016exceptionally, puy2016modulation, vecchi2006reassessing}. Its strength is given by a state-dependent (multiplicative) noise that depends on the SSTa \citep{jin2007ensemble, bianucci2018estimate, chen2023rigorous}, where a warmer SSTa leads to stronger wind burst activities. As for the decadal variability in the state variables through $I$, it stems from the observation that since 1870, through several detailed El Ni\~no-type classification methods, the EP and CP El Ni\~no events are alternatively prevalent every 10 to 20 years \citep{yu2013identifying, dieppois2021enso}. This oscillation between EP-dominant and CP-dominant regimes indicates that the decadal variability plays an important role in the underlying dynamics, which is parameterized through a simple linear stochastic differential equation with multiplicative noise \eqref{eq:walker}, with no explicit dependence on the state variables in the faster time scales \citep{yang2021enso}. They, together with additional small Gaussian white noise $\sigma_u\dot{W}_u$, $\sigma_h\dot{W}_h$, $\sigma_C\dot{W}_C$, $\sigma_E\dot{W}_E$, characterize the irregularity and multiscale features of the ENSO complexity \citep{timmermann2018nino, fang2020brief}.

It has been shown in the original work \citep{chen2022multiscale} that the model can reproduce many observed properties of the ENSO statistics and ENSO diversity. In terms of statistics, the model can reconstruct the observed power spectrums in both the CP and EP regions. It perfectly recovers the climatological PDFs of the SSTa indices with their strong non-Gaussian statistics. It also captures the observed seasonal phase-locking features. As for the ENSO complexity with respect to spatiotemporal patterns, the model can reproduce roughly the same ratio of EP to CP events and the intensity of these events, including the amplitude and frequency of the extreme ones, as in observations. The model can also produce delayed super El Ni\~no and mixed CP-EP events. Furthermore, the model generates multi-year events with more multi-year La Ni\~na than multi-year El Ni\~no, consistent with observations \citep{fang2020contrasting}.

In the following, the non-dimensional form of the model in \eqref{conceptual_model} is utilized to compute the responses. In the non-dimensional form, the six state variables have comparable maximum amplitudes. This means the strength of the components in the 6-dimensional column vector describing the eigenvector of the quadratic form can be used to intuitively tell the most sensitive direction. Nevertheless, the results for the evolution of the statistics with or without the perturbation are shown in the dimensional form that facilitates the physical explanations.

\section{Results}\label{Sec:Results}
\subsection{Statistical response to the perturbations of initial values}
\begin{figure}[ht]
    \hspace*{-0cm}\includegraphics[width=1.0\textwidth]{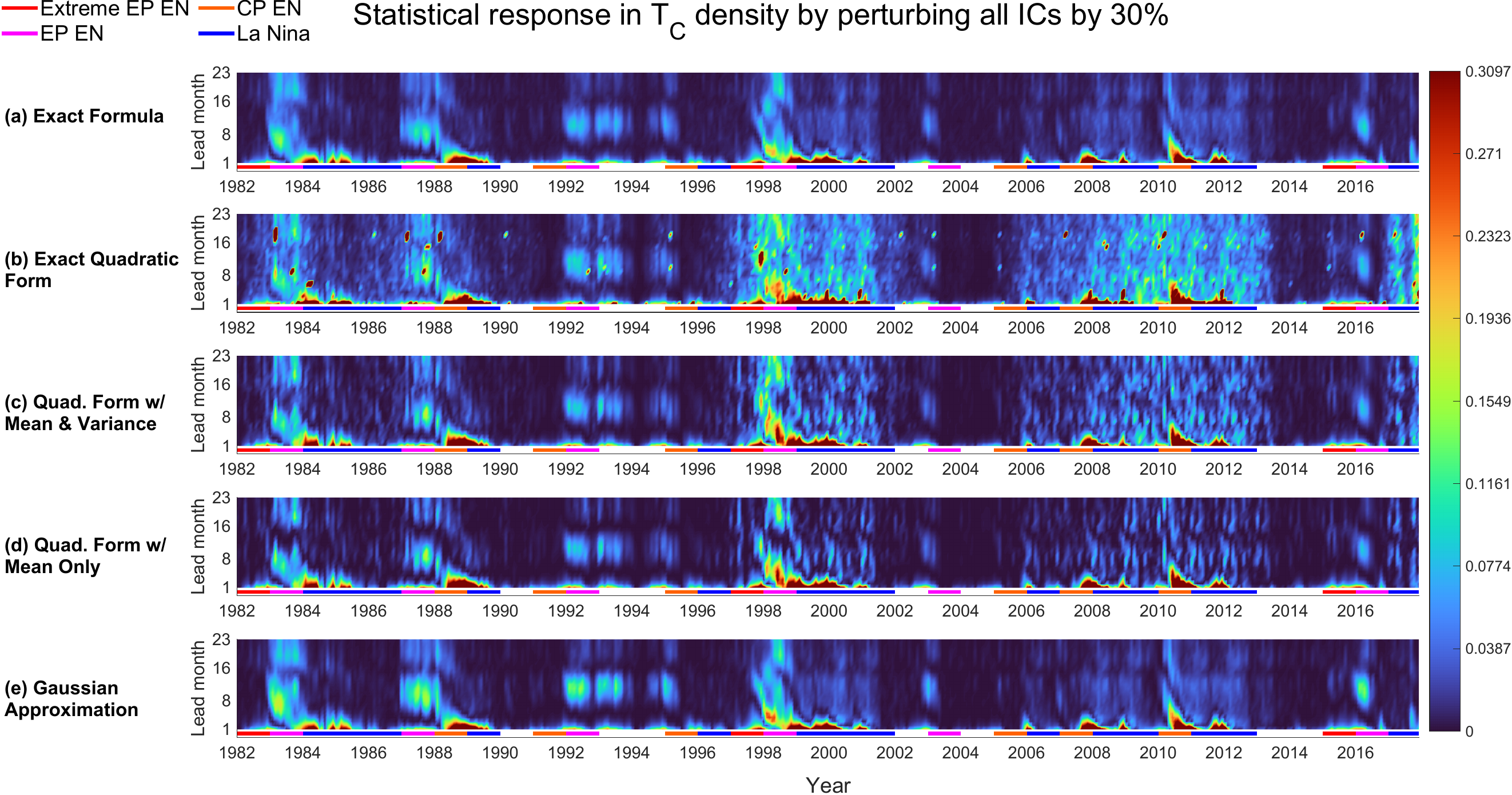}
    \caption{The statistical response of $T_C$ to the initial value perturbations for events with different starting dates and at different lead times, where a 30\% perturbation is added to the initial value of each of the six state variables. Different rows show the resulting statistical response amplitude measured by the relative entropy using different methods. In each row, the x-axis is the starting date on the first day of each month across the 36 years, and the y-axis is the lead time (months). The horizontal lines above the x-axis indicate the event type of that year based on the DJF SSTa.}
    \label{Response_comparison_T_C}
\end{figure}

Figure \ref{Response_comparison_T_C} shows the statistical response of $T_C$ to the initial value perturbations for events with distinct starting dates and at different lead times, where a 30\% perturbation is added to the initial value of each of the six state variables.

The following conclusions can be made from the result using the exact formula \eqref{relative_entropy} (Panel (a)). Overall, the statistical response at future time instants heavily relies on the initial state of the system. The initial perturbation has the most far-reaching impact on the subsequent evolution of the system when the initial phase is at the peak of strong EP events, where the response can remain considerable even after two years. This is unsurprising as a strong initial value can be easily amplified for a chaotic system as time goes on \citep{fang2023quantifying}. The responses from the extreme El Ni\~nos of 1982-1983 and 1997-1998 are the strongest among the different events. The initial perturbations of the other two significant El Ni\~nos, in 1987-1988 and 2015-2016, also have a long-range impact. The finding implies that the increased strength of El Ni\~no events under the climate change scenario not only affects the environment of those years but also has direct subsequent impacts over a long period. In contrast, if the initial state lies at a La Ni\~na event, the statistical response is only significant for a very short period. After that, the system follows the discharge mechanism, which is more predictable \citep{sharmila2023contrasting}. In such a case, the difference in the initial value, namely the perturbation, is often damped quickly. Furthermore, if the system is initially in a neutral state, then the amplitudes of different state variables are all near zero. Consequently, the perturbation, which is a percentage of the initial value, is insignificant and the subsequent response becomes negligible. Qualitatively similar results are found in $T_E$, although the response becomes more significant for $T_E$ when the perturbation is imposed at the initial phases of strong EP events. Finally, the phase-locking properties are preserved in the statistical response.

In addition to the general conclusion of the statistical response to initial perturbations, the intercomparison between different methods in Figure \ref{Response_comparison_T_C} provides the following two crucial findings. First, the statistical response using the Gaussian approximation (Panel (e)) leads to nearly the same results as that using the exact formula (Panel (a)). At first glance, this may look controversial as the Gaussian approximation cannot capture the strong non-Gaussian climatology statistics. However, despite missing the information in the higher-order moments, the Gaussian approximation may remain accurate in computing the relative entropy if the information gap between the two non-Gaussian distributions behaves similarly to that in the low-order Gaussian statistics. The comparable patterns in Panel (a) and Panel (e) indicate that the statistical response can be effectively computed using the much cheaper Gaussian approximation, which facilitates the use of more sophisticated models in practice. Second, the three approximations with the quadratic form (Panels (b)--(d)) overall lead to similar results as that using the exact formula. This is strong evidence indicating that the leading-order expansion of the relative entropy with the Fisher information in \eqref{quadratic_form}--\eqref{Fisher_Info} is appropriate in computing the statistical response. Yet, the exact quadratic form (Panel (b)), which uses the full PDF in computing the gradients \eqref{Fisher_Info}, gives noisier patterns. This is because taking the numerical gradient of the full PDF can be sensitive to small errors in the tail of the estimated distribution that affect the accuracy of the numerical method in Section \ref{Subsec:Summary}. Therefore, an appropriate practical strategy involves utilizing the low-order moments, e.g., the Gaussian statistics, as approximations. This becomes especially helpful when seeking the most sensitive perturbation directions, as shown below. It is worth noticing that the quad form w/ mean only gives similar response patterns as the exact formula within a short time range (less than 6 months). It results in more significant errors at long lead times and for those years with larger uncertainties (e.g., 1983-1984, 1987-1989, etc). Such a result indicates that the mean response is the dominant component in the total response to initial value perturbations for short lead times. This justifies using the standard trajectory-wise method in studying the response. As the lead time increases, some differences can be found between these two methods at medium-range lead times, which implies that the variance and higher-order moments take over the role of accounting for the statistical response. The following case studies will demonstrate such a feature. Finally, the statistical response decays to zero at a longer period as the chaotic system only has a finite memory length. Note that the findings here are very different from the scenario with the parameter perturbation (see below), where the response in the change of climatology is primarily attributed to the variance and high-order moments.

\begin{figure}[ht]
    \hspace*{-0cm}\includegraphics[width=1.0\textwidth]{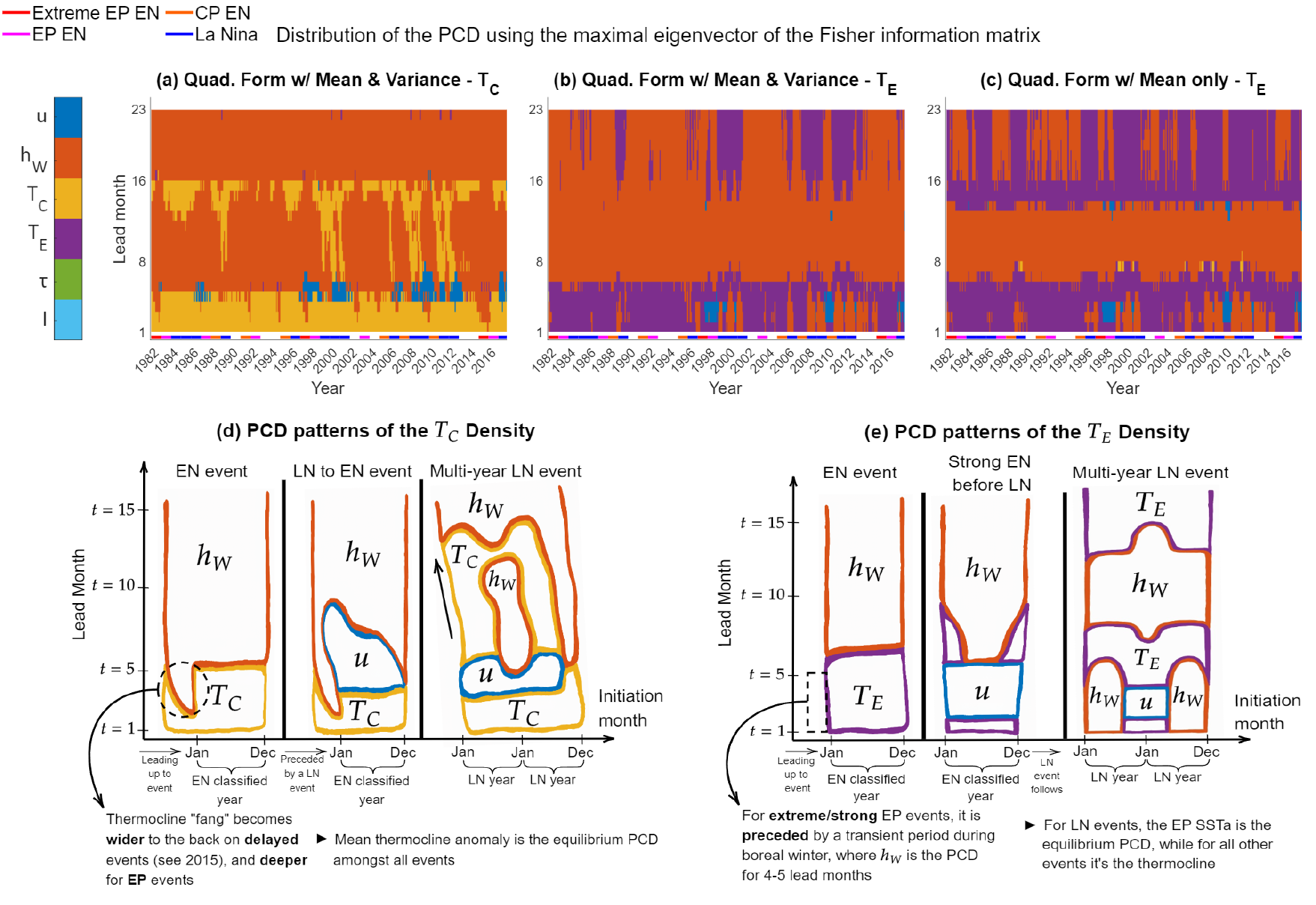}
    \caption{The most sensitive direction of perturbation. In Panels (a)--(b), the directions are computed using the quad form w/ mean and variance. Each point in the plot represents the variable associated with the largest component in the 6-dimensional eigenvector, namely the principal coordinate direction (PCD). It approximates the most sensitive perturbation direction if a perturbation is imposed on the corresponding starting date (its x-axis value) that leads to the response at a given lead time (its y-axis value). The two panels show the cases when the statistics of $T_C$ and $T_E$ are adopted, respectively, in computing the relative entropy. Panel (c) shows the PCD based on the statistical response of $T_E$ using the quad form w/ mean only. Panels (d)--(e) summarize the schematic structures of the most sensitive direction for different ENSO events corresponding to the findings in Panels (a)--(b).  }
    \label{MSD_IC_comparison}
\end{figure}

Figure \ref{MSD_IC_comparison} utilizes the principal coordinate direction (PCD) to demonstrate the most sensitive direction of perturbation at different lead times computed from the quad form w/ mean and variance. Note that the exact direction is given by a $6\times1$ vector consisting of the components for the state variables $(u,h_W, T_C, T_E,\tau, I)$. Yet, for the convenience of presentation, only the most significant component (in absolute value) in this 6-dimensional vector is used. This is named PCD. Later, the full eigenvector will be used to describe the most sensitive direction in the case studies. It will be seen in the case studies that several variables may all have non-negligible contributions at lead times ranging from 3 to 15 months. Typically, $T_E$ and $h_W$ will both contribute to the response of $T_E$ while $T_C$, $h_W$, $T_E$ and $u$ will all impact the response of $T_C$. The exact percentages of the contributions from each variable vary for different events. As was mentioned at the end of Section \ref{Subsec:Model}, the calculation is based on the non-dimensional system where the amplitudes of all the state variables are roughly the same. Thus, the largest component in this vector indeed reflects the dominant direction. Panels (a)--(b) show the PCD for all events across the 37-year observational period at different lead times when the statistical responses of the $T_C$ and $T_E$ are evaluated, respectively. Panels (d)--(e) summarize the schematic structures of the PCD for different events. Despite some inter-event differences, the overall patterns of each type of ENSO event are similar. As shown in Panel (d), perturbing $T_C$ always gives itself the strongest response at a short lead time, usually within three months. Yet, before the El Ni\~no events, especially the strong EP events, thermocline depth becomes the predominant component that triggers the strongest response of $T_C$. A stronger thermocline depth strengthens the recharge mechanism and changes the SST patterns, affecting the EP and the CP regions. The direction along the thermocline depth is also the choice for the initial perturbation to maximize the response in the future at the interannual time scale. Next, if the perturbation is imposed when a La Ni\~na event transits to an El Ni\~no one, the zonal advection can play an essential role in the response at around five months lead time. This is unsurprising as the advection helps accelerate the recharge process and modify the patterns at such a time scale \citep{an1999role, tao2023impacts}. The most complicated scenario is the multi-year La Ni\~na events, for example, 1999-2000, where $T_C$ and $h_W$ alternate as the most sensible variable from 5 to 15 months lead time, depending on when the perturbation is imposed. In contrast, as is shown in Panel (e), when the statistical response of $T_E$ is considered before an EP El Ni\~no event, perturbing $T_E$ gives the strongest response of itself for a short lead time. The PCD becomes $h_W$ as lead time increases. This simple structure is changed at the phase when an EP El Ni\~no transits to a La Ni\~na, where the advection again becomes important around a lead time of 5 months. For multi-year La Nina events, $T_E$ and $h_W$ alternate at different starting months as the most crucial variable to perturb that triggers the strongest response of $T_E$. Finally, Panel (c) shows the PCD based on the statistical response of $T_E$ but using the quad form w/ mean only method. The patterns in Panels (b) and (c) are similar for lead times of less than 6 months. Some differences can be seen around 7 to 9 months. The difference becomes more significant for longer ranges after 14 months, though the amplitude of the response is negligible since the initial effect almost goes away at such a long range.

\begin{figure}[ht]
    \hspace*{-0cm}\includegraphics[width=1.0\textwidth]{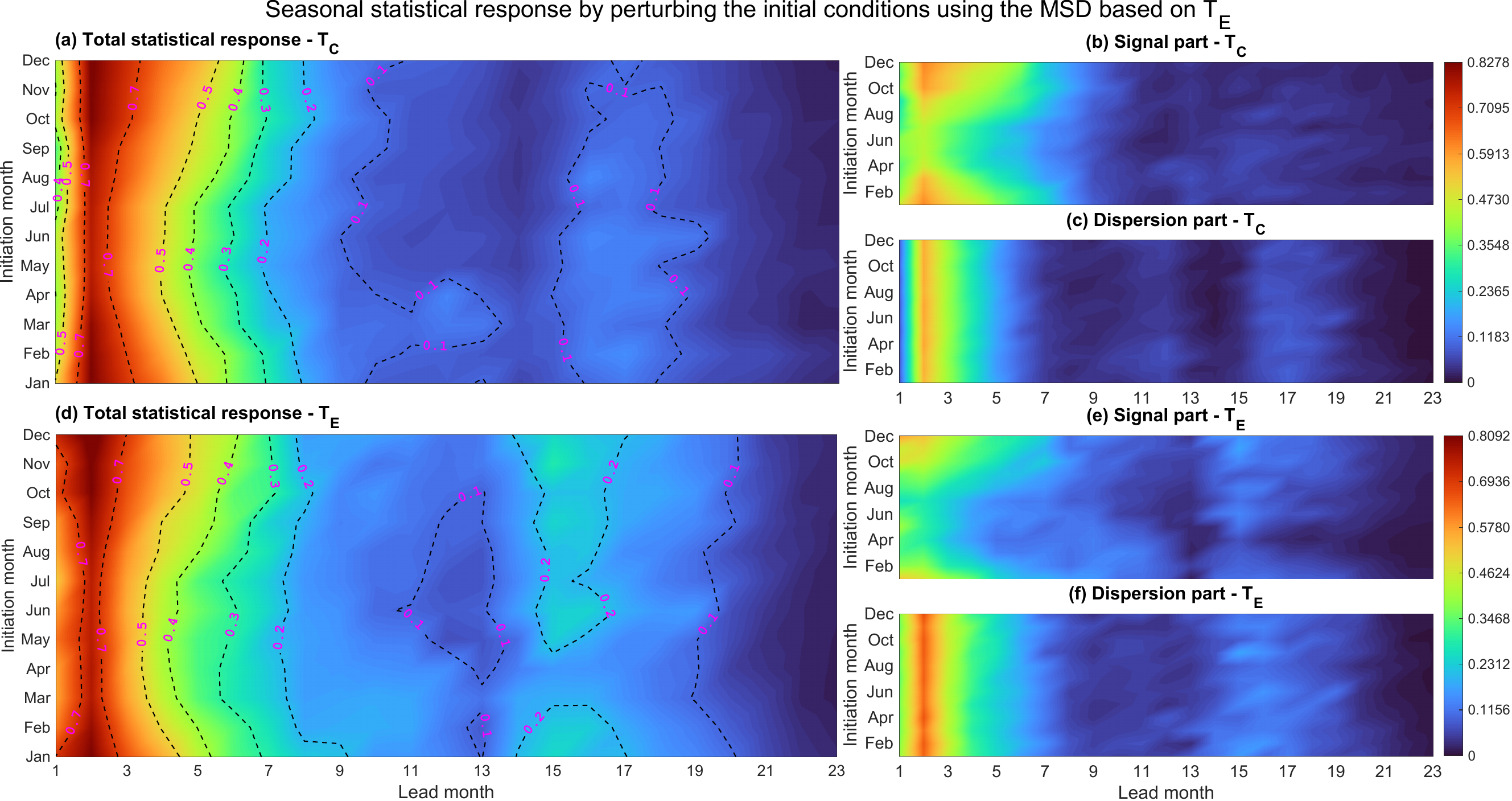}
    \caption{Seasonal statistical response by perturbing the initial conditions using the most sensitive direction (MSD) based on $T_E$. Panels (a)--(c): The response of $T_C$, including the total response and the response in the signal and the dispersion, respectively. Panels (d)--(f): The response of $T_E$.   }
    \label{SPB_study_RE_T_E_MSD_IC}
\end{figure}

\begin{figure}[ht]
    \hspace*{-0cm}\includegraphics[width=1.0\textwidth]{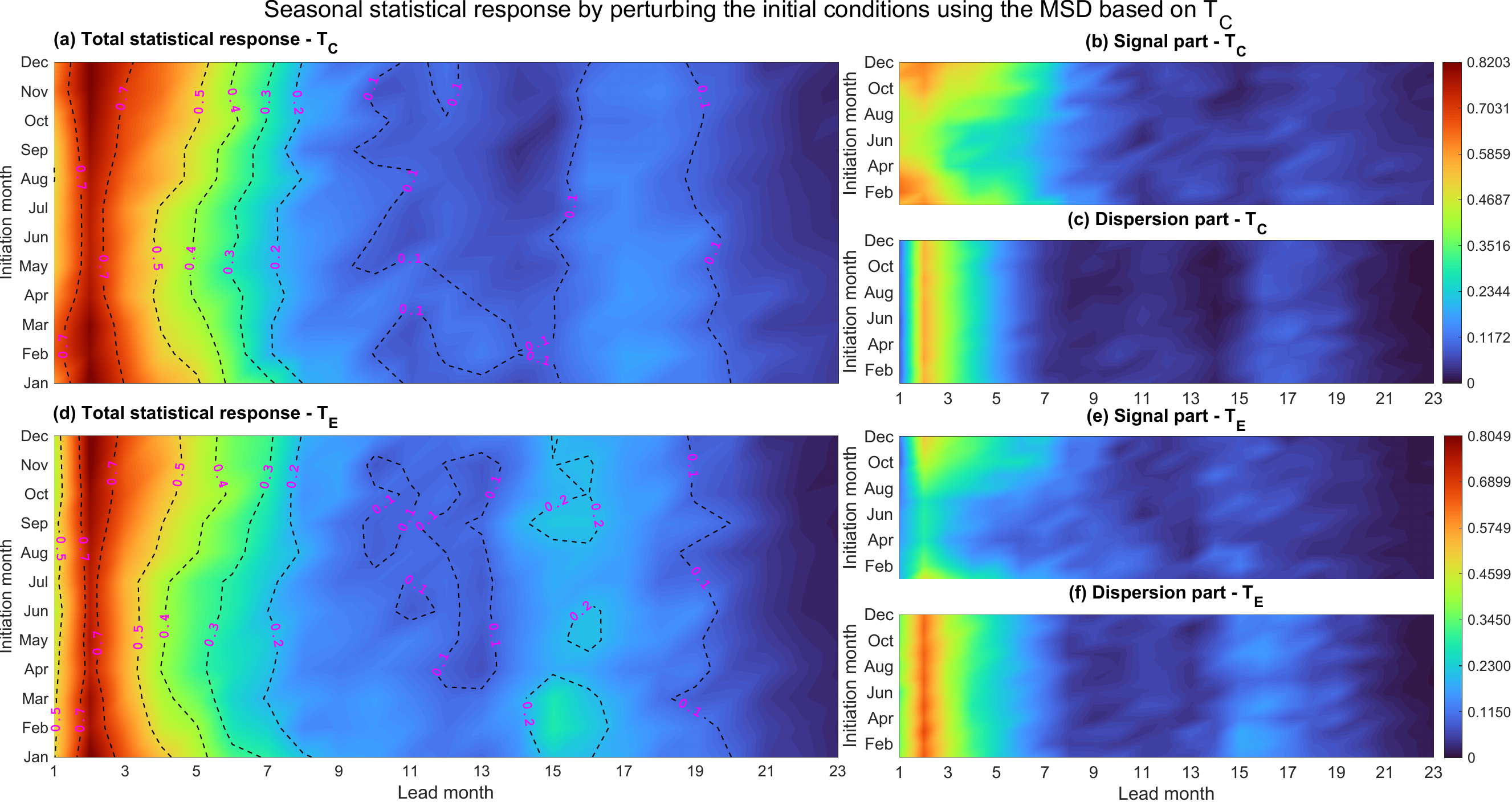}
    \caption{Similar as Figure \ref{SPB_study_RE_T_E_MSD_IC} but using the most sensitive direction (MSD) based on $T_C$.  }
    \label{SPB_study_RE_T_C_MSD_IC}
\end{figure}

Next, Figures \ref{SPB_study_RE_T_E_MSD_IC}--\ref{SPB_study_RE_T_C_MSD_IC} present the seasonal statistical response by perturbing the initial conditions along the most sensitive direction of either $T_E$ (Figure \ref{SPB_study_RE_T_E_MSD_IC}) or $T_C$ (Figure \ref{SPB_study_RE_T_C_MSD_IC}). Using Figure \ref{SPB_study_RE_T_E_MSD_IC} as an example, the procedure for generating such a figure is as follows. First, the statistical response is computed for different dates and lead times. This will give a plot similar to that in Panel (a) of Figure \ref{Response_comparison_T_C}. The difference compared with Figure \ref{Response_comparison_T_C} is that in Figure \ref{SPB_study_RE_T_E_MSD_IC}, the most sensitive direction for $T_E$ at the $k$-month lead is used to determine the initial perturbation direction when computing the response PDF at such a lead time. In other words, starting from the same date, different initial perturbations are adopted to calculate the response at different lead times. Then, the relative entropy between the response PDF and the unperturbed one is calculated. Finally, the relative entropy values are averaged over the dates with the same initial month and lead time to reach the plots. Panels (a) and (d) in these two figures show that similar to the spring prediction barrier \citep{lopez2014wwbs, duan2013spring, zheng2010spring}, there is a spring barrier for the response. Such a spring barrier is consistent when adding the initial perturbation based on the most sensitive direction of $T_E$ or $T_C$. The spring barrier is related to the overall weak initial strength of the signal in the boreal spring. However, compared with the spring barrier for the standard trajectory-wise prediction, the spring barrier for the statistical response is less significant. To understand such a difference, Panels (b)--(c) and (e)--(f) show the response corresponding to the signal and dispersion components, respectively. It is seen that the spring barrier is significant in the signal part. Note that the signal part, defined in \eqref{Signal_Dispersion}, is based on the mean time series, which can be regarded as a surrogate of a trajectory and is thus more consistent with the standard trajectory-wise-based spring prediction barrier. On the other hand, the dispersion part, which is based on the variance, shows no apparent spring barrier. This weakens the overall spring barrier in the statistical response. The insignificant spring barrier in the variance response is possibly due to the weak interaction between the mean and variance, so the time inhomogeneous behavior in the mean response does not affect the variance too much. The findings here indicate that the model response can behave differently in different statistical measurements regarding the spring barrier. The spring barrier affects the path-wise prediction but may not significantly influence the propagation of the overall uncertainty.

\clearpage
\subsection{Statistical response to the perturbations of model parameters}
\begin{figure}[ht]
\centering
    \hspace*{-0cm}\includegraphics[width=0.93\textwidth]{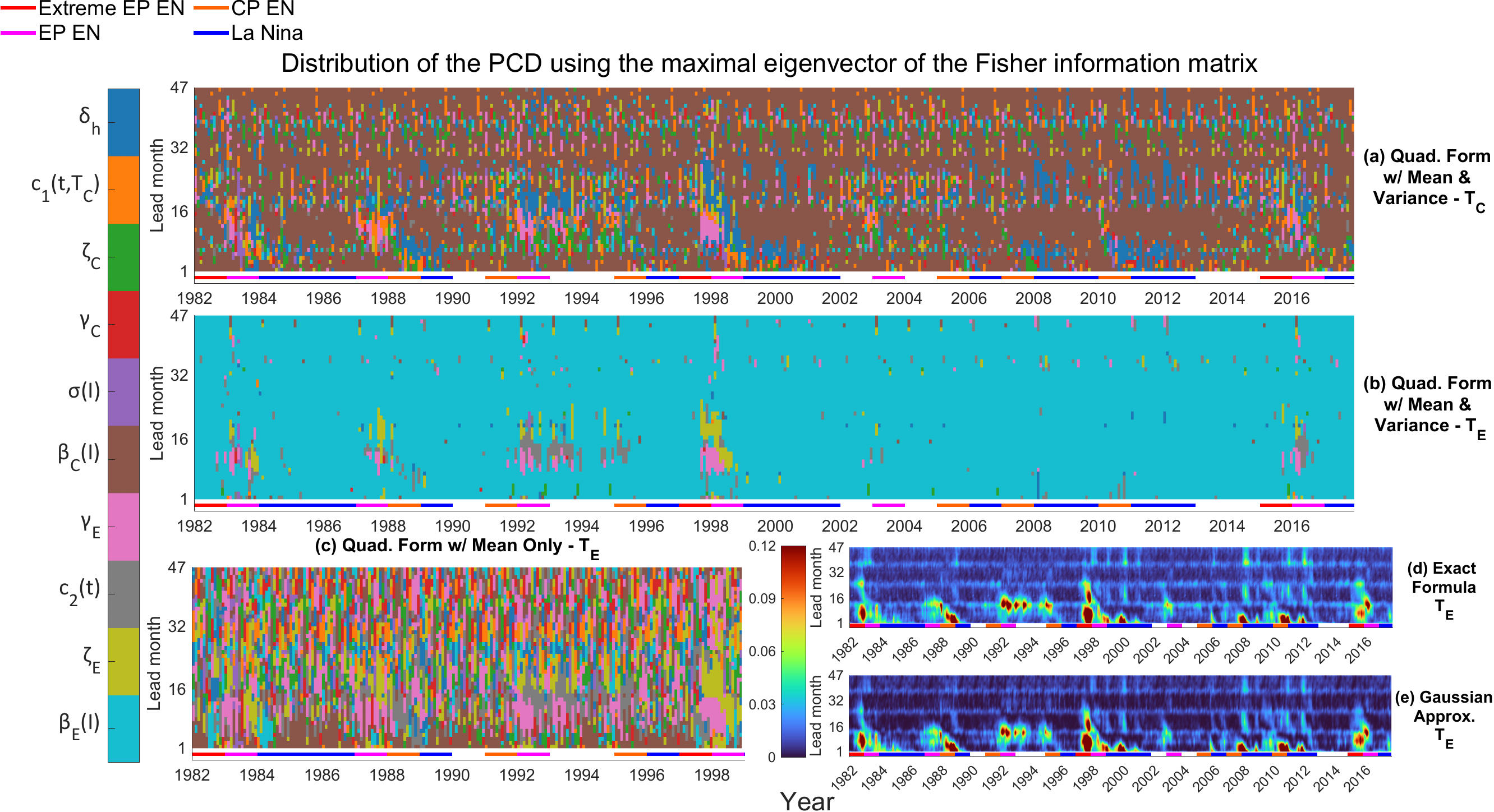}
    \caption{The most sensitive direction (in the form of the PCD) and the statistical response to the perturbations of model parameters. Panels (a)--(b) show the most sensitive directions for the response in $T_C$ and $T_E$, respectively, using the quad form w/ mean and variance. Panel (c) shows that in $T_E$ using the quad form w/ mean only. Panels (d)--(e) show statistical response of $T_E$ by perturbing all the variables by 10\% using the exact formula and the Gaussian approximation.}
    \label{MSD_Param_comparison}
\end{figure}

Figure \ref{MSD_Param_comparison} shows the PCD and the statistical response to the perturbations of model parameters. Note that when the parameters are perturbed, they will remain unchanged afterwards. Thus, the long-term statistics (i.e., the climatology distribution) will become different from the unperturbed system. This is distinct from the initial value perturbation that lasts only for a finite period due to the finite memory of the system.

The vector containing the parameters for perturbation is $10\times 1$ dimension. The following parameters are considered:
\begin{enumerate}
    \item $\delta_h$: The average SSTa feedback in the thermocline depth equation.
    \item $c_1(t,T_C)$: The damping coefficient in the $T_C$ equation, representing the nonlinear parametrization of the subsurface structure as well as the discharge behavior of the SSTa in the CP region.
    \item $\zeta_C$: The feedback coefficient of $T_E$ in the $T_C$ equation.
    \item $\gamma_C$: The thermocline feedback coefficient in the $T_C$ equation.
    \item $\sigma(I)$: The decadal variability coupling parameter (zonal ocean current coupling coefficient).
    \item $\beta_C(I)$: The wind stress coefficient in the $T_C$ equation; also related to the increase or decrease of the MJO or the tropical cyclones in the CP region.
    \item $\gamma_E$: The thermocline feedback coefficient in the $T_E$ equation.
    \item $c_2(t)$: The damping coefficient in the $T_E$ equation, representing the nonlinear parametrization of the subsurface structure as well as the discharge behavior of the SSTa in the EP region.
    \item $\zeta_E$: The feedback coefficient of $T_C$ in the $T_E$ equation.
    \item $\beta_E(I)$: The wind stress coefficient in $T_E$ dynamics; also related to the increase or decrease of the MJO or tropical cyclones in the EP region.
\end{enumerate}

It is seen from Panels (a)--(b) that the wind burst coefficients $\beta_C$ and $\beta_E$ are the dominant parameter that leads to the strongest response in the statistics of $T_C$ and $T_E$, respectively. These results are as expected, since strengthening the wind activities will significantly increase the SST amplitude and change the resulting statistics. The coefficient $\gamma_E$ in front of the thermocline depth $h_W$ and the coefficient $\zeta_E$ in front of the CP SSTa $T_C$ in the equation of $T_E$ are also shown to be important intermittently in affecting the statistical response of $T_E$. The parameter $\zeta_E$ and the coefficient $\delta_h$ in front of the SST feedback in the $h_W$ equation both affect the response of $T_C$. One notable finding by comparing Panel (b) and Panel (c) is that the most sensitive perturbation direction using the quad form w/ mean only gives a significantly different result than that using the quad form w/ mean and variance. It reveals that the uncertainty, reflected by the variance, plays a crucial role in determining the statistical response. The strengthening of the wind activity may not change the mean response but will significantly increase the variance of the response. Subsequently, it enhances the probability of the occurrence of extreme events. Section \ref{Subsec:CaseStudy} will include more detailed studies. Finally, as shown in Panels (d)--(e), Similar to the initial value perturbation, the calculated responses using the exact formula and the Gaussian approximation are similar, which again justifies utilizing the latter to improve the computational efficiency in practice. 
\subsection{Case studies}\label{Subsec:CaseStudy}
\subsubsection{Response to initial value perturbations for different ENSO events}
\begin{figure}[ht]
    \hspace*{-0cm}\includegraphics[width=1.0\textwidth]{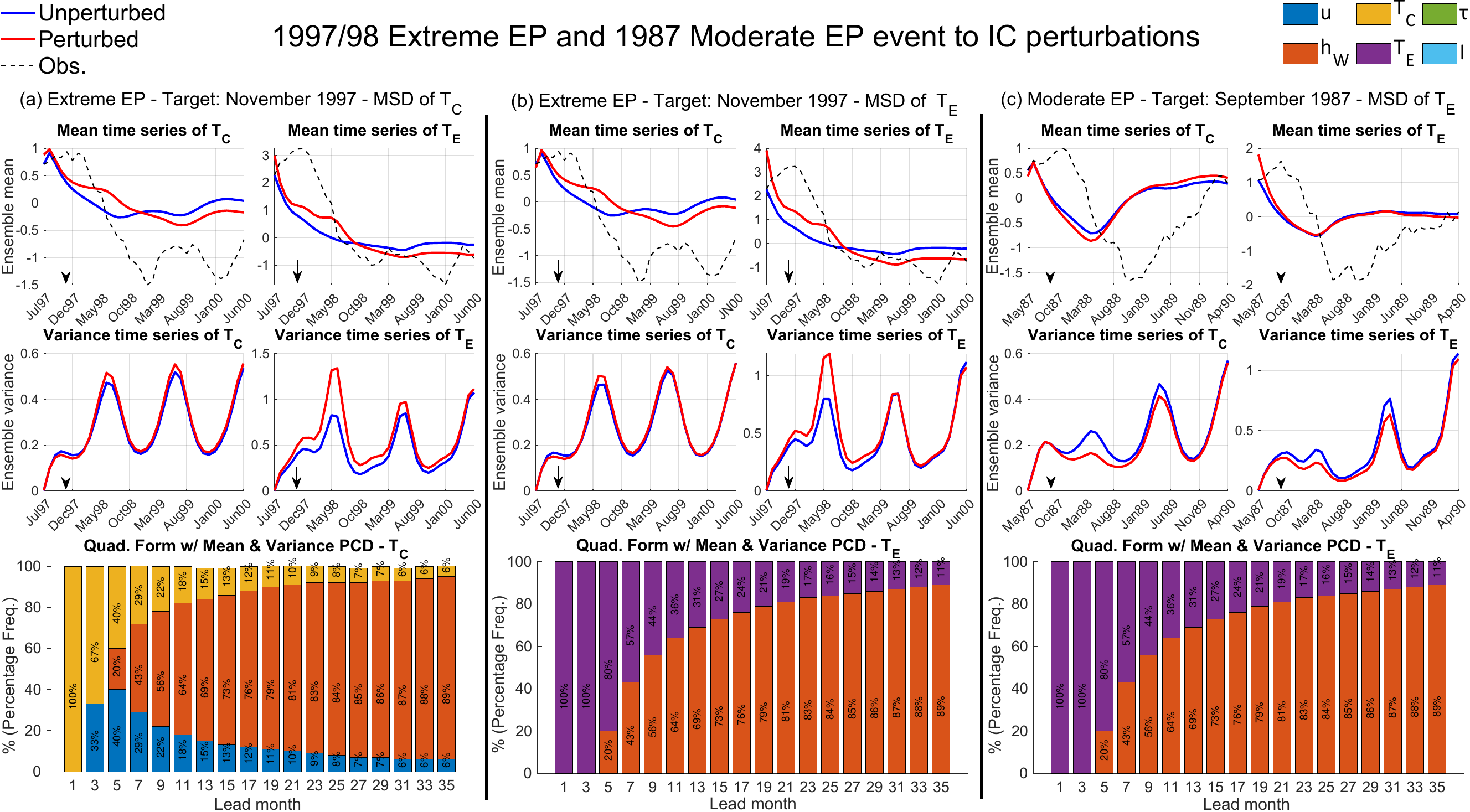}
    \caption{Statistical responses of the 1997 extreme EP El Ni\~no event (Columns (a)--(b)) and the 1987 moderate EP El Ni\~no event (Column (c)). The first two rows show the time evolution of the mean and variance. The starting date is 4 months before the event peak, where the event peak is marked by a black arrow. The perturbation at the initial time corresponds to the one that triggers the most sensitive perturbation at a lead time of 4 months, namely at the event peak, for $T_C$ (Column (a)) and $T_E$ (Columns (b)--(c)), respectively. The blue and red curves show the time evolution of the statistics associated with the unperturbed and the perturbed initial conditions, respectively. The dashed black curve in the panels of the mean time series shows the single true trajectory. The third row shows the most sensitive perturbation direction at different lead times, where the starting date is the same as the first two rows. }
    \label{Extreme_Moderate_EP}
\end{figure}

\begin{figure}[ht]
    \hspace*{-0cm}\includegraphics[width=1.0\textwidth]{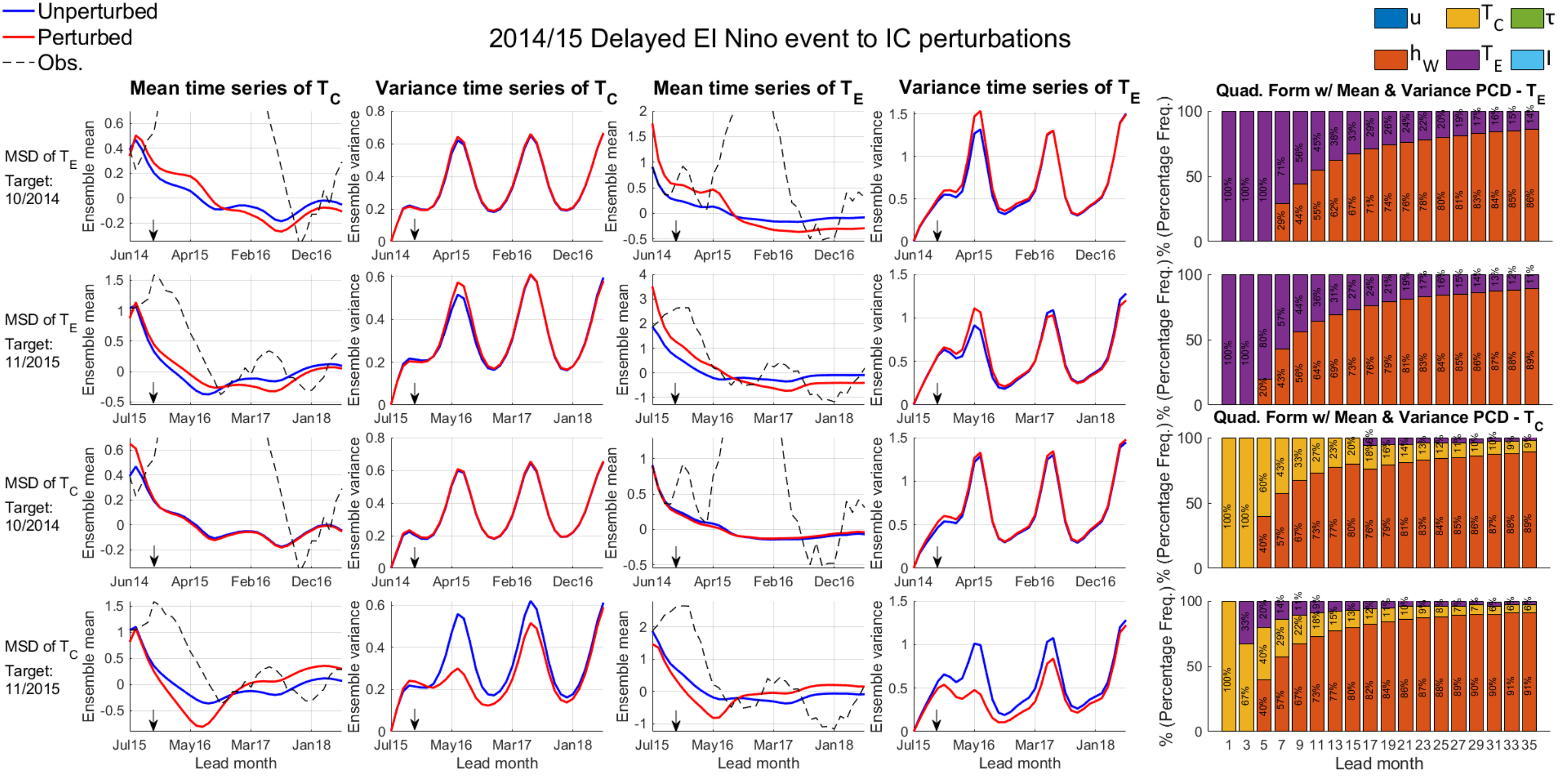}
    \caption{Statistical responses of the 2014-2015 delayed El Ni\~no event. The first four columns show the time evolution of the mean of $T_C$, variance of $T_C$, mean of $T_E$, and variance of $T_E$, respectively. The last column shows the PCD (as a surrogate for the most sensitive perturbation direction) at different lead times, where the starting date is the same as the corresponding first four rows. The first and the third rows show the results when the initial perturbation is imposed on June 2014, 4 months before the first peak of the event. The second and the fourth rows show the results when the initial perturbation is imposed on July 2015, 4 months before the second (and the strong) peak of the El Ni\~no event. The initial perturbation for the time series in the first two rows corresponds to the strongest response at a lead time of 4 months using $T_E$ while that in the last two rows uses the response of $T_C$. }
    \label{Delayed_EP}
\end{figure}

\begin{figure}[ht]
    \hspace*{-0cm}\includegraphics[width=1.0\textwidth]{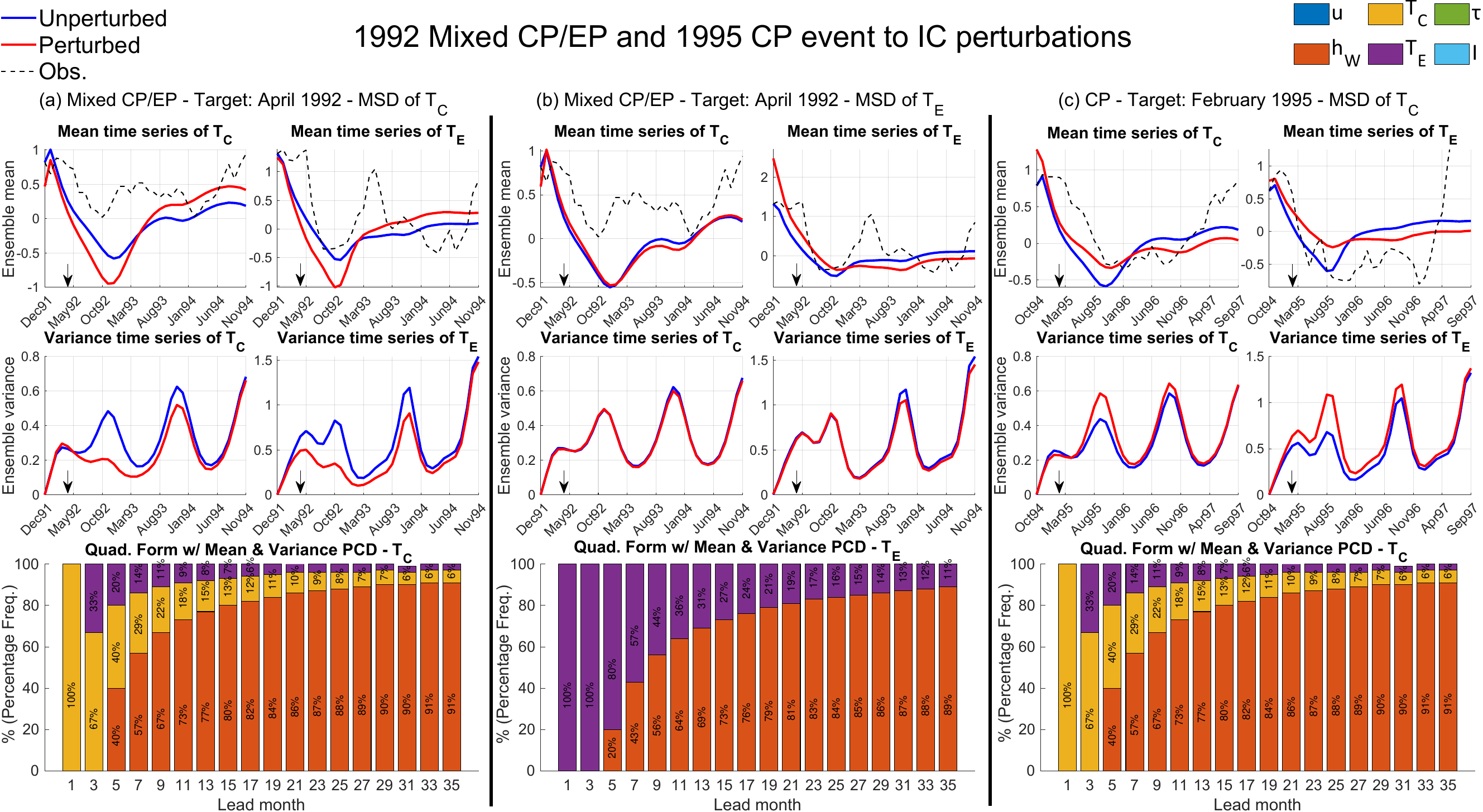}
    \caption{Statistical responses of the 1992 mixed CP-EP event (Columns (a)--(b)) and the 1995 CP El Ni\~no event (Column (c)). The caption description is similar to Figure \ref{Extreme_Moderate_EP}. }
    \label{Mixed_CP_EP_and_CP}
\end{figure}

\begin{figure}[ht]
    \hspace*{-0cm}\includegraphics[width=1.0\textwidth]{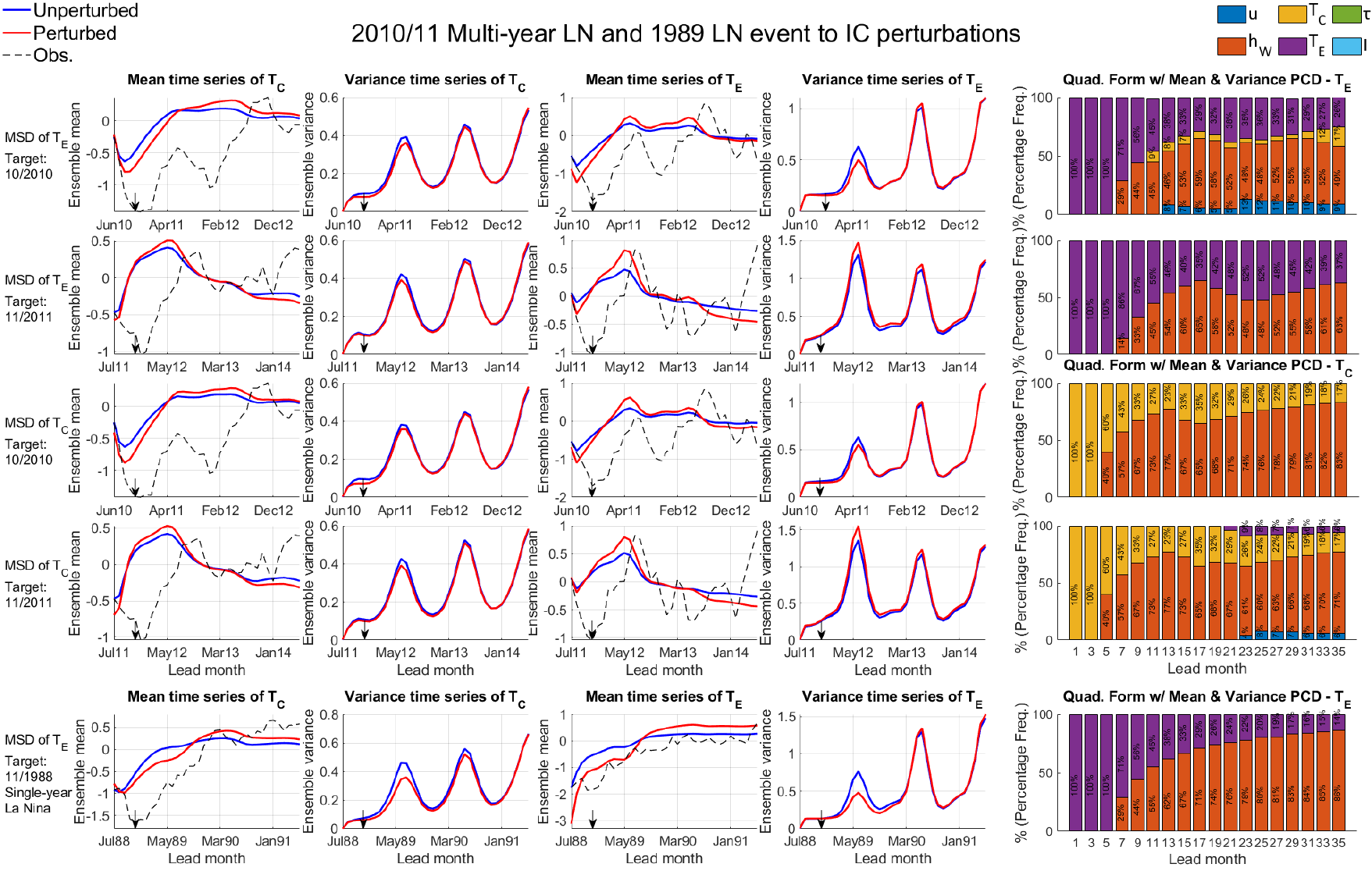}
    \caption{Statistical responses of the 2010-2011 multi-year La Ni\~na event and the 1988 single-year La Ni\~na event. The caption description is similar to Figure \ref{Delayed_EP}.}
    \label{MY_and_single_LN}
\end{figure}

Figure \ref{Extreme_Moderate_EP} shows the statistical response of the 1997 extreme EP El Ni\~no event (Columns (a)--(b)) and the 1987 moderate EP El Ni\~no event (Column (c)). For the time evolution of the mean and variance responses shown in the first two rows, the starting date is 4 months before the event peak. The perturbation at the initial time corresponds to the one that triggers the most sensitive perturbation at a lead time of 4 months, namely at the event peak, for $T_C$ (Column (a)) and $T_E$ (Columns (b)--(c)), respectively. The associated coordinate of the maximal eigenvector is used as the perturbation added to the initial condition. The third row of this figure shows the most sensitive perturbation direction at different lead times, where the lead time of 4 months corresponds to the event peak. Similar representations of the results are adopted in Figures \ref{Delayed_EP}, \ref{Mixed_CP_EP_and_CP}, and \ref{MY_and_single_LN}. Note that the results are robust within a certain range to the choice of the initial perturbations in terms of the lead time.

For the 1997 extreme event, perturbing $T_C$ and $T_E$ results in the strongest responses for themselves at a short lead time, respectively. This is as expected since the SST variables do not have time to respond to the perturbation of other variables within such a short time. At long lead times, the most sensitive direction is predominated by the thermocline feedback, which is the crucial variable for amplifying the discharge-recharge mechanism \citep{jin1999thermocline}. The zonal advection $u$ plays a vital role in the medium-range lead time from 3 to 9 months when considering the statistical response of $T_C$. It does not significantly impact the statistical response of $T_E$. This is consistent with a recent finding in \citep{tao2023impacts}, which suggests the zonal current error has the most substantial impact on the western and central tropical Pacific. Regardless of whether $T_C$ or $T_E$ is used to assign the initial perturbation, the time evolution of the mean and the variance responses are similar, as is seen in the first two rows of Columns (a) and (b). In both cases, the mean response dominates the time evolution of the statistical response for $T_C$. Starting from July 1997, the mean response is also the main contributor to the statistical response of $T_E$ for a short lead time, but the variance becomes equally essential in the total explained response from January to May 1998. A stronger response in the variance indicates a higher probability of triggering extreme events, which is never seen in the mean response time series that goes toward the neutral state. The 1987 moderate EP El Ni\~no event has a similar profile of the statistical response. The only difference is that the mean response quickly disappears, but the variance response lasts much longer until the follow-up La Ni\~na event. This means the perturbation of the precursor of the El Ni\~no event can have a far-reaching impact on the subsequent years and affect the associated discharge phase of the cold event.

Figure \ref{Delayed_EP} shows the statistical response of the 2014-2015 delayed El Ni\~no event \citep{allan2020placing, ludescher2014very, santoso2017defining}. The initial perturbation is imposed on June 2014, 4 months before the first peak of the event (first and third rows), and on July 2015, 4 months before the second and the strong peak of the El Ni\~no event (second and fourth rows). The initial perturbation for the time series in the first two rows corresponds to the strongest response at a lead time of 4 months using $T_E$ while that in the last two rows uses the response of $T_C$. The following conclusions can be made by comparing the responses in the first and the third rows. Starting from June 2014, along the most sensitive perturbation direction of $T_E$ (first row), the response in the mean of $T_C$ and $T_E$ dominates the total response while the response in the variance is negligible. In contrast, the overall statistical response is relatively weak along the most sensitive perturbation direction of $T_C$ (third row). Next, it can be seen from the second and the fourth rows that the situation becomes very different when the starting date becomes July 2015. Both the mean and the variance will contribute to the statistical response regardless of using $T_C$ or $T_E$ as the variable for determining the most sensitive perturbation direction. Notably, when the direction causing the strongest response of $T_C$ at the lead time of 4 months is utilized as the initial perturbation direction (the fourth row), the time evolution of the statistical response is significant for both the SST variables, especially in the variance response. This is similar to the 1997 extreme event shown in Figure \ref{Extreme_Moderate_EP}. The results imply that their statistical response remains the same despite the intrinsic differences in the formation mechanisms of these two strong El Ni\~no events. It is also worth remarking that wind bursts are crucial for each realization of the events \citep{chen2015strong, hu2016exceptionally, levine2016july, chen2017formation, thual2019statistical, thual2016simple}, not only in observations but also in the model utilized here. Nevertheless, the wind burst activity does not appear to trigger strong statistical responses of the SST variables. This is because the intraseasonal wind bursts occur in a much faster time scale than the SST variables. Therefore, their statistical contribution is averaged out. This also illustrates a fundamental difference in studying the El Ni\~no events using trajectory-wise and statistical methods.

Figure \ref{Mixed_CP_EP_and_CP} includes the cases of the 1992 mixed CP-EP event (Columns (a)--(b)) and the 1995 CP El Ni\~no event (Column (c)). For the mixed CP-EP event, the first two rows in Column (a) show the time evolution of the mean and variance for the unperturbed and perturbed initial conditions. The initial perturbation corresponds to the direction of the strongest response of $T_C$ at the lead time of 4 months. With the first 5 months, the mean response of $T_C$ outweighs the variance response while the response in $T_E$ is negligible. However, at the range of the 6- to 14-month lead times, both the mean and variance responses become significant for $T_C$ and $T_E$. Note that the unperturbed system predicts a 50\% possibility of a La Ni\~na event in October 1992 (since the mean value is $-0.5^o$C), though the true signal does not exceed this threshold. With the perturbation, the statistical forecast indicates that a La Ni\~na event will almost surely occur. Thus, such an initial perturbation is critical since it drives the dynamics from a possibly neutral state to a La Ni\~na event at a lead time of around one year. Column (b) shows that if the initial perturbation is imposed based on the strongest response of $T_E$ at the lead time of 4 months, then the perturbation has almost no influence on the statistical evolution of $T_C$ and the dominant component of the response in $T_E$ is in the mean. The comparison between these two columns implies the role of different initial perturbations in modulating the dynamics. In addition, the mean and variance responses are not always synchronized, especially for long lead times, which further indicates the importance of utilizing the entire PDF or at least considering the variance in calculating the response. Finally, by imposing the initial perturbation along the direction of the strongest response of $T_C$ at the lead time of 4 months, the time evolution of the statistical response of the 1995 CP El Ni\~no in Column (c) is similar to the CP-EP event in Column (a). That is, both the mean and the variance responses are significant for lead times up to 14 months.

\begin{figure}[ht]
    \hspace*{-0cm}\includegraphics[width=1.0\textwidth]{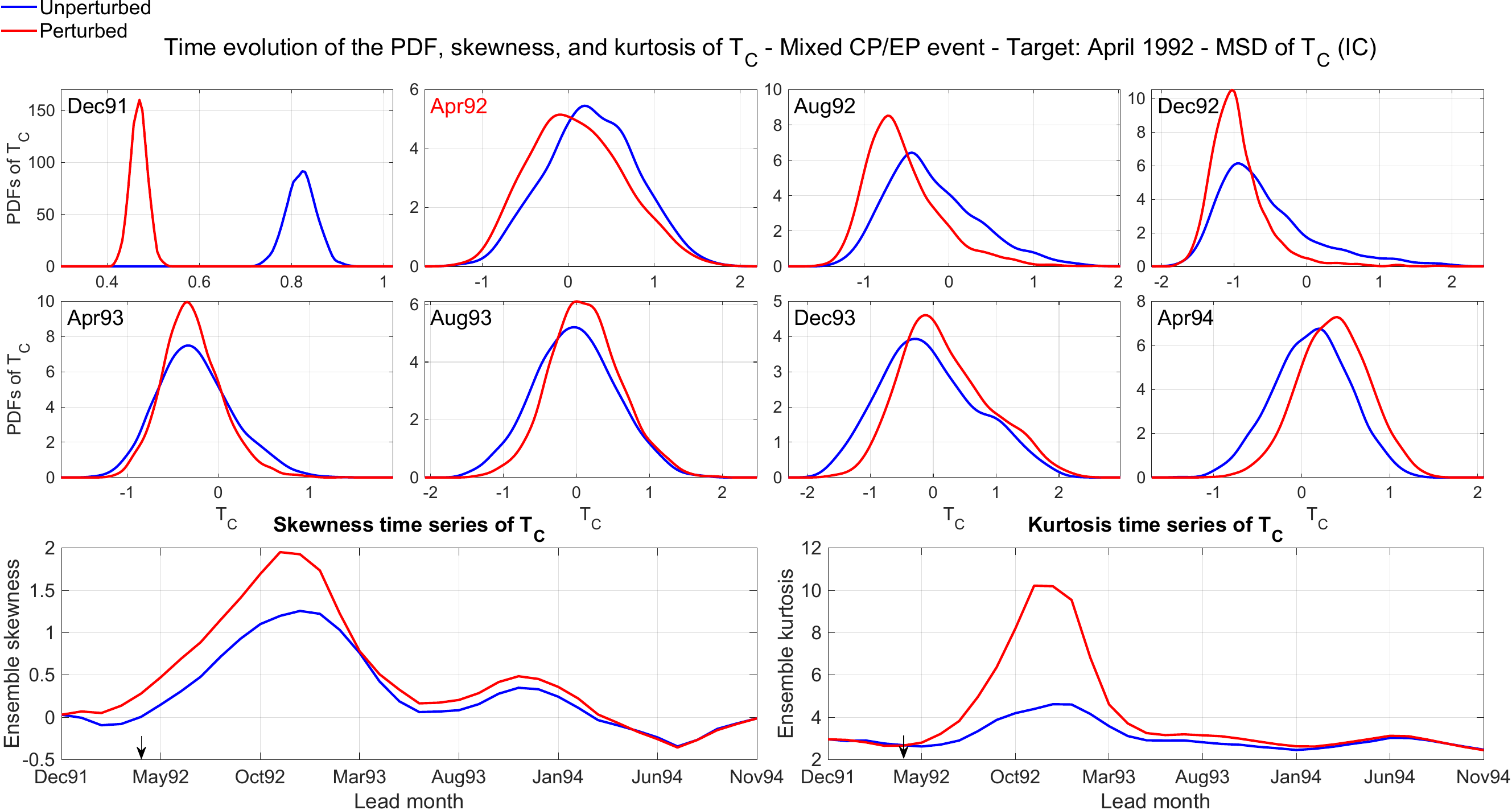}
    \caption{Time evolution of the statistics of $T_C$ of the 1992 mixed CP-EP event when the perturbation is imposed on December 1991. The initial perturbation corresponds to the direction of the strongest response of $T_C$ at the lead time of 4 months. This figure supplements Column (a) of Figure \ref{Mixed_CP_EP_and_CP}. The first two rows show the time evolution of the PDFs associated with the unperturbed (blue) and perturbed (red) initial conditions. The third row shows the time evolution of the third and fourth moments, namely the skewness and kurtosis, representing the non-Gaussian features. }
    \label{Mixed_CP_EP_PDF_skew_kurt}
\end{figure}

Figure \ref{Mixed_CP_EP_PDF_skew_kurt} demonstrates the time evolution of the entire PDF and that of the skewness and kurtosis of $T_C$ of the 1992 mixed CP-EP event. The initial perturbation corresponds to the direction of the strongest response of $T_C$ at the lead time of 4 months. This figure supplements Column (a) of Figure \ref{Mixed_CP_EP_and_CP}. A strong response in the PDF of $T_C$ happens at the end of 1992, which is consistent with that in the leading two moments in Figure \ref{Mixed_CP_EP_and_CP}. Notably, the PDF associated with the perturbed initial condition can show even more non-Gaussian features than the one with the unperturbed initial condition. Nevertheless, since the skewness and kurtosis peak at the same time as the variance, the result justifies that the Gaussian approximation is appropriate for finding the most sensitive perturbation direction.

Finally, Figure \ref{MY_and_single_LN} shows the case studies of the 2010-2011 multi-year La Ni\~na event \citep{iwakiri2021mechanisms, luo2017inter} and the 1988 single-year La Ni\~na event. For the 2010-2011 multi-year La Ni\~na event, initial perturbations are imposed on two different dates, October 2010 and November 2011. The two dates are 4 months before the peak of the first and second year La Ni\~na. For the single-year La Ni\~na event, the perturbation is imposed 4 months before the peak time. As in the previous figures, the initial perturbation corresponds to the direction of the strongest response of $T_C$ or $T_E$ at the lead time of 4 months. For these La Ni\~na events, the mean response is the dominant component of the statistical response. This indicates that the initial perturbation is dissipated over the discharge phase, which explains the longer predictability of the La Ni\~na events.

\clearpage

\subsubsection{Response to model parameter perturbations}
\begin{figure}[ht]
    \hspace*{-0cm}\includegraphics[width=1.0\textwidth]{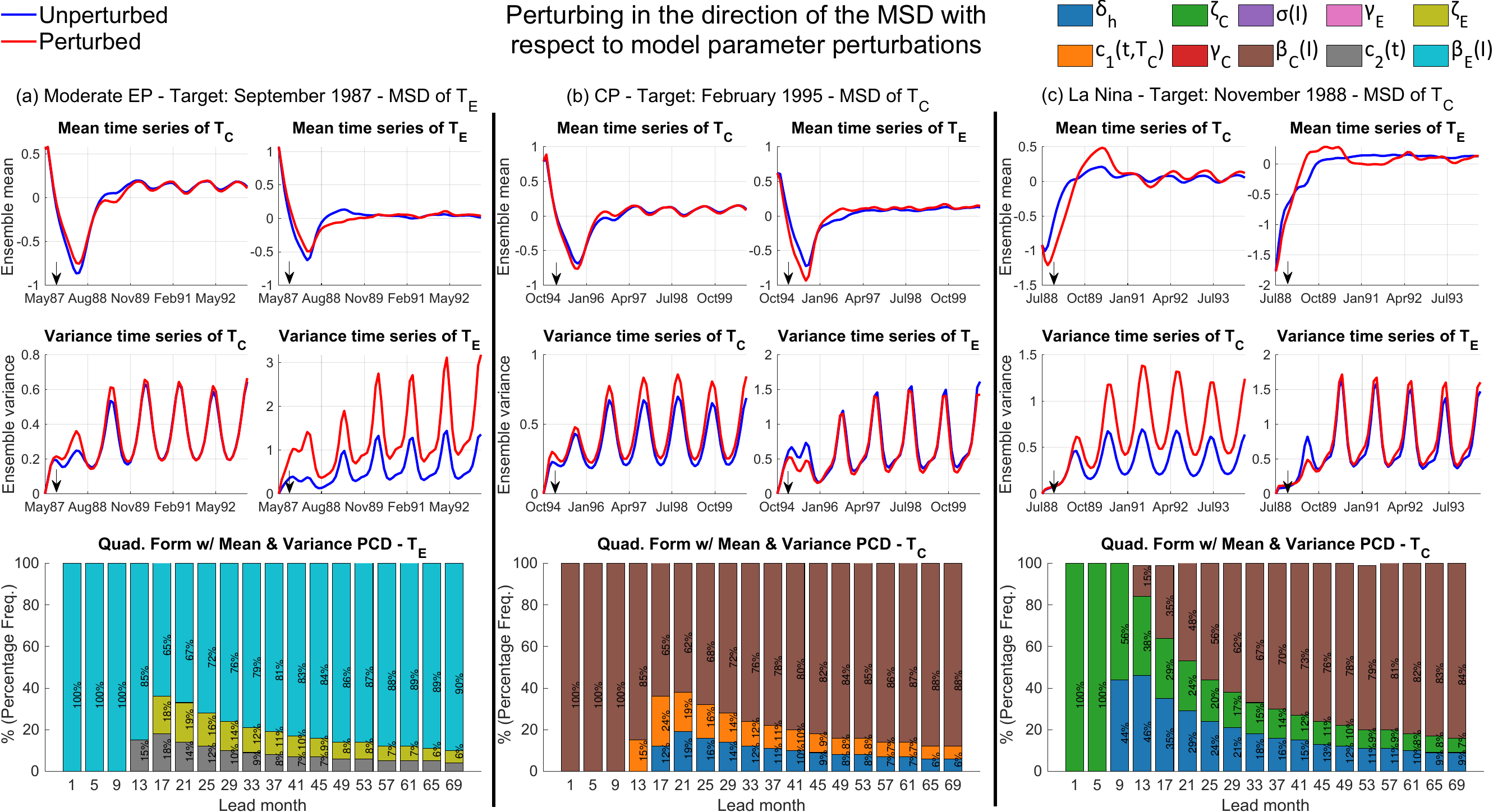}
    \caption{Statistical response to parameter perturbations. The first two rows show the time evolutions of the mean and variance using the unperturbed (blue) and perturbed (red) initial condition. The perturbation is imposed 4 months ahead of the event peak. The direction of the perturbation is determined by triggering the largest perturbation of either $T_E$ (Column (a)) or $T_C$ (Columns (b)--(c)) at the event peak using the quad form w/ mean and variance method. The third row shows the PCD at different lead months starting 4 months ahead of the event peak. Columns (a)--(c) show the cases of the 1987 moderate EP El Ni\~no, the 1995 CP El Ni\~no, and the 1988 La Ni\~na events, respectively.}
    \label{MP_EP_CP_LN}
\end{figure}

Figure \ref{MP_EP_CP_LN} shows three case studies about the statistical response to model parameter perturbations. In all cases, perturbation is imposed 4 months before the event's peak. The direction of the perturbation is determined by triggering the largest perturbation of either $T_E$ (Column (a)) or $T_C$ (Columns (b)--(c)) at the event peak using the quad form w/ mean and variance method.  

Column (a) shows the case of the 1987 moderate EP El Ni\~no. The most sensitive perturbation direction in terms of the statistics of $T_E$ at the lead time of 4 months completely follows the direction of perturbing the wind stress coefficient $\beta_E$ in the $T_E$ equation. Since the wind stress coefficient only changes the amplitude of the wind but not the preferences of the WWB or the EWB, the mean response is zero. However, because of the strengthening of the wind forcing, the response in the variance of $T_E$ significantly increases. Consequently, the probability of the occurrence of extreme events becomes large. Note that despite the coupling between $T_C$ and $T_E$, the resulting variance response of $T_C$ is only significant for about a year after the initial perturbation.

Column (b) shows the case of the 1995 CP El Ni\~no, where the most sensitive perturbation direction is determined by maximizing the statistical response of $T_C$ at the lead time of 4 months. As a symmetry to Column (a), the most sensitive direction is entirely given by the wind stress coefficient $\beta_C$ in the $T_C$ equation. Again, the perturbed statistics are mainly due to the change of the variance rather than the mean, which triggers more variabilities in the subsequent SST.

Column (c) shows the case of the 1998 La Ni\~na. Since the wind burst is very weak during the La Ni\~na phase. The wind stress coefficient is no longer the major contributor to a strong statistical response. The perturbation of the parameter $\zeta_C$, representing the feedback from $T_E$ in the $T_C$ equation, is the one that triggers the strongest response in the dynamics of $T_C$ in the medium range within 4 to 7 months. In contrast, the feedback coefficient $\delta_h$ accounting for the feedback from $T_C$ and $T_E$ to the dynamics of $h_W$ is the one that triggers the most significant response for the 8 to 13 months lead time. Finally, the wind stress coefficient $\beta_C$ becomes the dominant factor for a longer range after the system leaves the La Ni\~na phase.

It is worth highlighting that when computing the most sensitive perturbations with parameter perturbations, the response in the variance is much more significant than that in the mean. A strong response in the variance and higher-order statistical moments implies the increased probability of extreme events and the enhancement of the uncertainty from the model output. This indicates the fundamental difference between statistical and traditional trajectory-wise methods in studying the model response and seeking the most sensitive direction. It is also remarkable that the parameter perturbation will change the climatology. Notably, when the difference mainly lies in the variance and higher-order moments, the response can be insignificant when applying the trajectory-wise methods. This indicates the necessity of using the statistical response computed from the information theory.

Finally, Table \ref{Table:Occurrence_frequency_ERSSTv5} shows the occurrence frequency of different ENSO events in observations (first row), original system (second row), and the perturbed systems by perturbing each listed parameter by 30\% (third to seventh rows). The observational period is from 1950 to 2020, totaling 71 years. The results here indicate the responses of the climatology to the change of these parameters. When perturbing $\beta_E$ and $\beta_C$, the EP and CP El Ni\~no events will become more frequent. Note that the number of extreme EP events is significantly increased when the wind stress or the wind amplitude is amplified, corresponding to an enhanced $\beta_E$. It also triggers more multi-year La Ni\~na events. Increased occurrences of the strong El Ni\~no and multi-year La Ni\~na are related to the global warming scenario \citep{geng2023increased, cai2015increased}. Next, when perturbing $\zeta_C$, the feedback coefficient of $T_E$ in the $T_C$ equation, the system tends to increase the number of CP events and decrease the number of EP events, pushing more activities toward the CP region due to the strengthening of the feedback. A decrease of $\delta_h$, the average SSTa feedback in the thermocline depth equation, will increase the chance of multi-year El Ni\~no events. Finally, recall that $c_1$ is the damping coefficient in the TC equation, representing the nonlinear parametrization of the subsurface structure as well as the discharge behavior of the SSTa in the CP region. If such a parameter is decreased, then the number of CP events will dramatically increase due to the more active response in the CP region \citep{zhao2021breakdown}. It will also trigger more multi-year El Ni\~no events but fewer multi-year La Ni\~na events. We would like to mention here that similar results are obtained when instead of the ERSSTv5 data we use the GODAS dataset to calculate the number of ENSO events over the 36-year observational period of 1982-2017 (where in this case 60 non-overlapping segments are used, with each one being 36 years long). The choice of the longer observational period provided by the ERSSTv5 dataset rests solely on the fact that the results are more robust in the statistical sense, due to the minimization of any biases that may occur when using a shorter time period.

\begin{table}[ht]{\footnotesize
\begin{tabular}{|lccccccc|}
\hline
\multicolumn{8}{|l|}{\textbf{ENSO Events Occurrence Frequency}}                                                                                                     \\ \hline
\multicolumn{1}{|l|}{}                        & \multicolumn{1}{c|}{El Ni\~no (EN)} & \multicolumn{1}{c|}{EP EN}         & \multicolumn{1}{c|}{CP EN}         & \multicolumn{1}{c|}{Extreme EP}  & \multicolumn{1}{c|}{Multi-year EN} & \multicolumn{1}{c|}{La Ni\~na (LN)} & Multi-year LN \\ \hline
\multicolumn{1}{|l|}{Observations} & \multicolumn{1}{c|}{24}           & \multicolumn{1}{c|}{14}           & \multicolumn{1}{c|}{10}           & \multicolumn{1}{c|}{4}           & \multicolumn{1}{c|}{5}             & \multicolumn{1}{c|}{24}           & 8             \\ \hline
\multicolumn{1}{|l|}{Original system}  & \multicolumn{1}{c|}{22.6$\pm$2.8} & \multicolumn{1}{c|}{13.4$\pm$2.4} & \multicolumn{1}{c|}{\phantom{0}9.2$\pm$2.1}  & \multicolumn{1}{c|}{4.5$\pm$1.7} & \multicolumn{1}{c|}{4.0$\pm$1.5}     & \multicolumn{1}{c|}{29.9$\pm$3.0}   & \phantom{0}7.5$\pm$1.8   \\ \hline
\multicolumn{1}{|l|}{Perturbing $\beta_E(I)$ $(\uparrow)$} & \multicolumn{1}{c|}{24.4$\pm$2.8} & \multicolumn{1}{c|}{17.0$\pm$2.9}   & \multicolumn{1}{c|}{\phantom{0}7.4$\pm$2.7}  & \multicolumn{1}{c|}{9.3$\pm$2.4} & \multicolumn{1}{c|}{5.4$\pm$1.5}   & \multicolumn{1}{c|}{33.8$\pm$3.0}   & 10.0$\pm$2.1    \\ \hline
\multicolumn{1}{|l|}{Perturbing $\beta_C(I)$ $(\uparrow)$} & \multicolumn{1}{c|}{22.2$\pm$3.3} & \multicolumn{1}{c|}{12.2$\pm$2.4} & \multicolumn{1}{c|}{10.0$\pm$2.6}   & \multicolumn{1}{c|}{4.3$\pm$2.0}   & \multicolumn{1}{c|}{3.8$\pm$1.9}   & \multicolumn{1}{c|}{31.0$\pm$3.0}     & \phantom{0}8.2$\pm$2.3   \\ \hline
\multicolumn{1}{|l|}{Perturbing $\zeta_C$ $(\uparrow)$}    & \multicolumn{1}{c|}{22.3$\pm$2.8} & \multicolumn{1}{c|}{11.6$\pm$1.8} & \multicolumn{1}{c|}{10.7$\pm$1.9} & \multicolumn{1}{c|}{4.1$\pm$1.6} & \multicolumn{1}{c|}{3.9$\pm$1.8}   & \multicolumn{1}{c|}{30.8$\pm$3.4} & \phantom{0}7.5$\pm$2.5   \\ \hline
\multicolumn{1}{|l|}{Perturbing $\delta_h$ $(\downarrow)$}   & \multicolumn{1}{c|}{22.5$\pm$3.0}   & \multicolumn{1}{c|}{13.7$\pm$2.9} & \multicolumn{1}{c|}{\phantom{0}8.8$\pm$2.8}  & \multicolumn{1}{c|}{4.4$\pm$2.1} & \multicolumn{1}{c|}{4.6$\pm$1.9}   & \multicolumn{1}{c|}{29.8$\pm$4.2} & \phantom{0}7.8$\pm$2.4   \\ \hline
\multicolumn{1}{|l|}{Perturbing $c_1(t,T_C)$ $(\downarrow)$} & \multicolumn{1}{c|}{26.8$\pm$2.9} & \multicolumn{1}{c|}{12.6$\pm$2.0}   & \multicolumn{1}{c|}{14.2$\pm$2.4} & \multicolumn{1}{c|}{4.1$\pm$1.8} & \multicolumn{1}{c|}{5.3$\pm$1.7}   & \multicolumn{1}{c|}{31.1$\pm$2.5} & \phantom{0}6.4$\pm$2.0     \\ \hline
\end{tabular}\caption{The occurrence frequency (i.e., the number of the events) of different ENSO events in observations (first row), original system (second row), and the perturbed systems by perturbing each listed parameter by 30\% (third to seventh rows). The observational period is from 1950 to 2020 (based on the data included in the ERSSTv5 dataset), totaling 71 years. 30 non-overlapping segments, each being 71 years long as to be consistent with the length of the observations, are simulated. The mean occurrence frequency numbers (per 71 years, i.e.\ over these 30 runs), plus and minus the corresponding standard errors of these segments, are shown. } \label{Table:Occurrence_frequency_ERSSTv5}}
\end{table}
\clearpage
\section{Conclusion}\label{Sec:Conclusion}
In this paper, a mathematical framework for computing the statistical response of a complex system is developed, where information theory is utilized to measure the strength of the response.  The method is then applied to study the response of different ENSO events to the perturbations of initial conditions and model parameters. It is also utilized to find the most sensitive perturbation direction for each ENSO event. The main conclusions are summarized as follows.
\begin{itemize}
  \item Depending on the initial phase and the time horizon, different state variables contribute to the most sensitive perturbation direction. While initial perturbations in SST and thermocline depth usually lead to the most significant response of SST at short- and long-range, respectively, initial adjustment of the zonal advection can be crucial to trigger strong statistical responses at medium-range around 5 to 7 months, especially at the transient phases between El Ni\~no and La Ni\~na.
  \item Despite the mean response dominating the total response with a short range for the initial value perturbation, the variance and higher-order moments contribute to the response at medium-range lead times.
  \item The spring barrier in the statistical response is overall weaker than that in the standard trajectory-wise prediction. Notably, the spring barrier is only significant in the signal part of the response PDF (corresponding to the mean) but is not apparent in the dispersion part (corresponding to the variance or uncertainty).
  \item The response in the variance triggered by external random forcing perturbations, such as the wind bursts, often dominates the mean response at long range in the parameter perturbation scenario, making the resulting most sensitive direction very different from the trajectory-wise methods.
  \item Despite the strong non-Gaussian climatology distributions, using Gaussian approximations in the information theory is efficient and accurate to compute the statistical response, allowing the method to be applied to more sophisticated systems, such as the intermediate coupled models in \citep{chen2023simple, geng2022enso} or operational systems.
\end{itemize}

It is worth emphasizing the different considerations when studying the two types of perturbations analyzed in this work: The initial conditions and the model parameters. The initial value perturbation will immediately impact each single ENSO event at short- and medium-range lead times. Since the role of the initial condition weakens as a function of the lead time, the long-term statistics are not affected by the initial value perturbation. In contrast, the parameter perturbation may not have a major impact at a short lead time. The response will become more significant as time evolves. It can also permanently change the climatology and is, therefore, more related to climate change. Despite this discrepancy in the goals, the response can be computed for both types of perturbations.

The study in this work highlights the response in the statistical sense. The most sensitive perturbation direction corresponds to the most influential perturbation that leads to the largest difference in the statistical forecast of the ENSO. Such difference is quantified using the information measurement, e.g., the relative entropy in Figure \ref{Response_comparison_T_C}. From a broader viewpoint, both the statistical method developed here and the trajectory-wise approaches aim to find the optimal precursors that trigger the most significant change in future states. However, the statistical framework differs from many existing methods based on computing the error in the trajectories, which do not highlight the role of the uncertainty, such as the variance. Since ENSO and many other natural phenomena are chaotic and contain uncertainty, it is essential to study the difference in the optimal precursors using a single trajectory and statistics. Understanding the types of events for which the most sensitive perturbation direction will strongly depend on the variance and the higher-order moments is also extremely helpful. The most sensitive perturbation direction can also be used to improve our understanding of ENSO physics. The variables that contribute the most to these significant perturbations can be explored to discover the triggering conditions of the corresponding event and understand the gap between events in the same category but with different strengths and amplitudes.

There are a few other topics that remain as potential future work. First, as Monte Carlo methods are needed to compute the statistics, the computational cost can be increased significantly when a large number of sample points is required for operational systems. The quadratic form provides a relatively cheap practice method by computing only the leading two moments. Overall, this study shows that the resulting response using such an approximation has a similar behavior to that of using the full distribution. Yet, a more rigorous quantification of the potential errors introduced in the approximation can be an interesting topic. Second, the implication of the statistical response on the statistical forecast can be further explored. Particularly, it is interesting to study the relationship between the long-term statistical response and the potential predictability of the ENSO. Third, the statistical response can be utilized in the multi-model scenario, potentially advancing the model selection and quantifying the model error.

\clearpage
\acknowledgments

The research of N.C. is funded by ARO W911NF-23-1-0118. M.A. is supported as a graduate research assistant under this grant.

%
%
\datastatement

The monthly ocean temperature and current data were downloaded from GODAS (https://www.esrl.noaa.gov/psd/data/gridded/data.godas.html). The daily zonal wind data at 850 hPa were downloaded from the NCEP–NCAR reanalysis (https://psl.noaa.gov/data/gridded/data.ncep.reanalysis.html). The Extended Reconstructed Sea Surface Temperature, Version 5 (ERSSTv5) data were downloaded from NOAA (https://psl.noaa.gov/data/gridded/data.noaa.ersst.v5.html).

The code used in the analysis and generation of the figures in this work can be provided upon contact with the corresponding author.

%

\appendix

\subsection*{Parameter values and characteristic scales}

In Table \ref{tbl:parameters} we include the model parameters in the standard run (i.e., unperturbed system), and the characteristic scales appearing in the three-region multiscale stochastic model \eqref{eq:adv}--\eqref{eq:walker}.

\begin{table*}[ht]
\begin{center}
\begin{tabular}{|l|clc|}
\hline
\multicolumn{3}{|c|}{$\varrho$ (scaling factor)} & \multicolumn{1}{c|}{$0.65$} \\
\hline
$b_0$        & \multicolumn{1}{c|}{$2.5$}          & \multicolumn{1}{l|}{$\mu$}          & $0.5$               \\
$\alpha_1$        & \multicolumn{1}{c|}{$0.0625\varrho^2=\alpha_2\varrho/2$}          & \multicolumn{1}{l|}{$\alpha_2$}          & $0.125\varrho$               \\
\hline
$[u]$        & \multicolumn{1}{c|}{$1.5$ms$^{-1}$}          & \multicolumn{1}{l|}{$[h]$}          & $150$m               \\
$[T]$        & \multicolumn{1}{c|}{$7.5^{\circ}$C}          & \multicolumn{1}{l|}{$[\tau]$}          & $5$ms$^{-1}$         \\
$[t]$        & \multicolumn{1}{c|}{$2$ months}              & \multicolumn{1}{l|}{$d_{\tau}$}        & $2$                  \\
$r$          & \multicolumn{1}{c|}{$0.25\varrho$}           & \multicolumn{1}{l|}{$r_C$}          & $0.75b_0\mu\varrho/2$      \\
$r_E$        & \multicolumn{1}{c|}{$3r_C=2.25b_0\mu\varrho/2$}   & \multicolumn{1}{l|}{$\lambda$}            & $2/60$               \\
$\delta_u$        & \multicolumn{1}{c|}{$\alpha_1b_0\mu$}       & \multicolumn{1}{l|}{$\delta_h$}          & $2\delta_u=\alpha_2b_0\mu$      \\
$\zeta_C$        & \multicolumn{1}{c|}{$0.75b_0\mu\varrho/2$}              & \multicolumn{1}{l|}{$\zeta_E$}          & $0.75b_0\mu\varrho/2$           \\
$\gamma_C$        & \multicolumn{1}{c|}{$0.75\varrho$}                 & \multicolumn{1}{l|}{$\gamma_E$}          & $0.75\varrho$              \\
$C_u$        & \multicolumn{1}{c|}{$0.03\varrho$}                 & \multicolumn{1}{l|}{$m$}            & $2$                  \\
$\sigma(I)$       & \multicolumn{1}{c|}{$I\varrho/5$}                  & \multicolumn{1}{l|}{$\beta_E(I)$}       & $0.15(2-0.2I)\sqrt{\varrho}$       \\
$\beta_u(I)$     & \multicolumn{1}{c|}{$-0.2\beta_E(I)$}            & \multicolumn{1}{l|}{$\beta_h(I)$}       & $-0.4\beta_E(I)$         \\
$\beta_C(I)$     & \multicolumn{1}{c|}{$0.8\beta_E(I)$}             & \multicolumn{1}{l|}{$\sigma_u$}          & $0.04\sqrt{\varrho}$               \\
$\sigma_h$        & \multicolumn{1}{c|}{$0.02\sqrt{\varrho}$}                  & \multicolumn{1}{l|}{$\sigma_C$}          & $0.04\sqrt{\varrho}$               \\
$\sigma_E$        & \multicolumn{1}{c|}{$0$}                     & \multicolumn{1}{l|}{$\sigma_I(I)$}       & $\sqrt{\lambda(4-I)I}$     \\
\hline
$\sigma_\tau(t,T_C)$ & \multicolumn{3}{c|}{$0.9[\tanh(7.5T_C)+1]\times\left[1+0.3\cos\left(\frac{2\pi}{6}t+\frac{2\pi}{6}\right)\right]$} \\
$c_1(t,T_C)$ & \multicolumn{3}{c|}{$\varrho\left[25\left(T_C+\frac{0.75}{7.5}\right)^2+0.9\right]\times\left[1+0.3\sin\left(\frac{2\pi}{6}t-\frac{\pi}{6}\right)\right]$}                                                                                     \\
$c_2(t)$     & \multicolumn{3}{c|}{$1.4\varrho\left[1+0.3\sin\left(\frac{2\pi}{6}t+\frac{2\pi}{6}\right)+0.25\sin\left(\frac{2\pi}{3}t+\frac{2\pi}{6}\right)\right]$}                                                                                     \\ \hline
\end{tabular}
\end{center}
\caption{Parameters of the stochastic conceptual model for the ENSO complexity \eqref{conceptual_model}.}
\label{tbl:parameters}
\end{table*}


\subsection*{Maximum entropy principle and coarse-grained statistical measurements}

When studying a stochastic dynamical system we are interested in some statistical quantities which we can measure through a family of $L$ functionals that represent different statistics of the dynamics through $\mathbf{x}_t$, which we denote by $\mathbf{E}_L(\mathbf{x}_t)=(E_1(\mathbf{x}_t),\ldots,E_L(\mathbf{x}_t))^\mathtt{\, T}$. At each lead time, we can extract $L$ measurements observed from the present dynamics or simulated by the model that correspond to our measurement functionals of interest, which we denote by $\overline{\mathbf{E}}_L=\left(\overline{E}_1,\ldots,\overline{E}_L\right)^\mathtt{\, T}$. This measured information of the dynamics acts as a restriction, with $\overline{\mathbf{E}}_L$ imposing $L$ constraints that are defined through the functionals $\mathbf{E}_L(\mathbf{x}_t)$, with each component defined as
\begin{equation*}
    \overline{E}_l=\int_{\mathbf{x}_t}E_l(\mathbf{x}_t)p^{\delta}(\mathbf{x}_t)\mathrm{d}\mathbf{x}_t, \ l=1,\ldots,L.
\end{equation*}
A natural choice for $E_l(\mathbf{x}_t)$ is the multivariate centralised moment of $\mathbf{x}_t$ of order $l=1,\ldots,L$ given by
\begin{equation*}
    \mu^{\delta}_{t,l}:=\begin{cases}        \int_{\mathbf{x}_t}\left\|\mathbf{x}_t\right\|p^{\delta}(\mathbf{x}_t)\mathrm{d}\mathbf{x}_t, & l=1 \\
    \int_{\mathbf{x}_t}\left\|\mathbf{x}_t-\overline{\mathbf{x}}_t\right\|^lp^{\delta}(\mathbf{x}_t)\mathrm{d}\mathbf{x}_t, & l\geq 2,
    \end{cases}
\end{equation*}
where $\overline{\mathbf{x}}_t=\mathbb{E}_{p}\left[\mathbf{x}_t\right]$ and the expectation is taken with respect to the (true) unperturbed dynamics. As such, $E_1(\mathbf{x}_t)=\left\|\mathbf{x}_t\right\|$ and $E_l(\mathbf{x}_t)=\left\|\mathbf{x}_t-\overline{\mathbf{x}}_t\right\|^l, l\geq 2$. In the main text we had $E_1(\mathbf{x}_t)=\mathbf{x}_t=\mathbf{x}_t^1$ and $E_l(\mathbf{x}_t)=\left(\mathbf{x}_t-\overline{\mathbf{x}}_t\right)^2$. This vector exponentiation can be equivalently interpreted as either the Euclidean norm raised to said power (as above), or as the product of the elements of the vector when applying that exponent element-wise. For either choice the quantities retrieved are equivalent, i.e.\ each measure can bound the other via a constant depending only on $l$, choice of norm, and $\mbox{Dim}(\mathbf{x}_t)$ \citep{stuart2010kendall}.

For $\pi$ being a PDF we define its differential entropy as
\begin{equation*}
    \mathcal{S}(\pi)=-\int_{\mathbf{x}}\pi(\mathbf{x})\log(\pi(\mathbf{x}))\mathrm{d}\mathbf{x}.
\end{equation*}
By empirical information theory \citep{jaynes1957information, majda2006nonlinear}, we can then construct at each lead time the unique least biased PDF under the Maximum Entropy Principle (MEP), $p^{\delta}_L$, which is the PDF that maximizes the differential entropy under the constraints of the $L$ observed measurements $\overline{\mathbf{E}}_L$. For general $L$ constraints imposed by $\overline{\mathbf{E}}_L$ and defined by the functionals $\mathbf{E}_L(\mathbf{x}_t)$, we have that the maximum entropy distribution is a member of the $L$-parameter exponential family of distributions \citep{majda2006nonlinear}, given as
\begin{equation*}
   p^{\delta}_L(\mathbf{x}_t)=e^{-\alpha_0-\boldsymbol{\alpha}_L\cdot \mathbf{E}_L(\mathbf{x}_t)},
\end{equation*}
where $\boldsymbol{\alpha}_L=\boldsymbol{\alpha}_L(t)=(\alpha_1(t),\ldots,\alpha_L(t))^T$ are the $L$ Lagrange multipliers chosen such that
\begin{equation*}
    \overline{\mathbf{E}}_L=\int_{\mathbf{x}_t}\mathbf{E}_L(\mathbf{x}_t)e^{-\alpha_0-\boldsymbol{\alpha}_L\cdot \mathbf{E}_L(\mathbf{x}_t)}\mathrm{d}\mathbf{x}_t,
\end{equation*}
while $\alpha_0=\alpha_0(t)$ is determined via the normalisation condition
\begin{equation*}
    e^{\alpha_0}=\int_{\mathbf{x}_t}e^{-\boldsymbol{\alpha}_L\cdot \mathbf{E}_L(\mathbf{x}_t)}\mathrm{d}\mathbf{x}_t.
\end{equation*}

Under the information theory-based framework, $\mathcal{P}(p^{\delta}(\mathbf{x}_t),p(\mathbf{x}_t))$ precisely quantifies the statistical response of a turbulent system due to the effects of a perturbation imposed on the initial state, internal fluctuations, or external forcings driving the dynamics. This quantification implicitly assumes the full statistical knowledge of the unperturbed dynamics and the availability of the perturbed densities. Such an assumption is unfeasible both practically and theoretically, thus making the need to use coarse-grained measurements, either observed or model-simulated ones, in conjunction with the MEP, necessary to construct the least-biased PDF under these constraints. The practical numerical approaches for computing the statistical response which were analyzed in section 2\ref{Subsec:PracticalApproaches} facilitate this procedure, specifically the method which utilizes the Fisher information matrix with coarse-grained statistical measurements. For the unperturbed dynamics, with true densitiy $p(\mathbf{x}_t)$, we can use $p_L(\mathbf{x}_t)$ as a surrogate, constructed by the MEP and the observational data from the unperturbed dynamics (known as the climatology), while for $p^{\delta}(\mathbf{x}_t)$ we can use the density constructed by the MEP and $L'$ model-averaged measurements from a perturbed simulation instead, where $L'\leq L$, which we denote by $p^{M,\delta}_{L'}(\mathbf{x}_t)$. This is because, in practice, the model density is determined by no more information than that available in the observations. For the statistical discrepancy between the true distributions and these computable and practical surrogates, measured via their relative entropy, it is possible to show that
\begin{equation*}
    \hspace{-0.24cm}\mathcal{P}[p^{\delta}(\mathbf{x}_t),p_{L'}^{M,\delta}(\mathbf{x}_t)]=\{\mathcal{S}[p_{L}^{\delta}(\mathbf{x}_t)]-\mathcal{S}[p^{\delta}(\mathbf{x}_t)]\}+\mathcal{P}[p_{L}^{\delta}(\mathbf{x}_t),p_{L'}^{M,\delta}(\mathbf{x}_t)]=a_0^M+\boldsymbol{\alpha}^{M}_{L'}\cdot \overline{\mathbf{E}}_{L'}-\mathcal{S}[p^{\delta}(\mathbf{x}_t)],
\end{equation*}
where $\boldsymbol{\alpha}^{M}_{L'}$ are the $L'$ model-determined Lagrange multipliers and $\alpha_0^M$ the respective normalisation constant from the MEP, and
\begin{equation*}
    \mathcal{P}[p(\mathbf{x}_t),p_{L}(\mathbf{x}_t)]=\mathcal{S}[p_{L}(\mathbf{x}_t)]-\mathcal{S}[p(\mathbf{x}_t)],
\end{equation*}
where, in particular, both convey the fact that the unbiased intrinsic error in the finite number of climate observations or model-averaged measurements, for both the perturbed and unperturbed climates, is exactly the entropy difference. These relations are what allow for the optimization principles of determining the model for which the true climate (perturbed or not) has the smallest additional information beyond the modeled climate distribution to be computably feasible \cite{majda2010quantifying}. A proof of these relations in the setting of predictability and model error can be found in p. 3-4 of \cite{majda2005information}.

We end this exposition by proving \eqref{quadratic_form_approx}. Since the observed statistics $\overline{\mathbf{E}}_L$ determine the perturbed MEP density $p^{\delta}_L$, with the Lagrange multipliers $\boldsymbol{\alpha}_L$ and normalisation constant $\alpha_0$, by differentiating the MEP density with respect to $\boldsymbol{\delta}$ we then have that
\begin{equation*}
    \cfrac{\left(\boldsymbol{\delta}\cdot\nabla_{\boldsymbol{\delta}}p_L^{\delta}(\mathbf{x}_t)\right)^2}{p_L^{\delta}(\mathbf{x}_t)}=\left((\boldsymbol{\delta}\cdot\nabla_{\boldsymbol{\delta}})\left(\alpha_0+ \boldsymbol{\alpha}_L\cdot \mathbf{E}_L(\mathbf{x}_t)\right) \right)^2p_L^{\delta}(\mathbf{x}_t),
\end{equation*}
where we use the fact that the family of measurement functionals in question does not depend on the perturbation (under some regularity assumptions). The gradients above are calculated at the unperturbed state of $\boldsymbol{\delta}=\boldsymbol{0}$. Further differentiating $\alpha_0$ with respect to $\boldsymbol{\delta}$ and using a bit of vector calculus yields
\begin{align*}
    \nabla_{\boldsymbol{\delta}}\alpha_0&=e^{-\alpha_0}\int_{\mathbf{x}_t}\nabla_{\boldsymbol{\delta}}\left(e^{-\boldsymbol{\alpha}_L\cdot \mathbf{E}_L(\mathbf{x}_t)}\right)\mathrm{d}\mathbf{x}_t \\
    &= \int_{\mathbf{x}_t}\nabla_{\boldsymbol{\delta}}\left(-\boldsymbol{\alpha}_L\cdot \mathbf{E}_L(\mathbf{x}_t)\right)e^{-\alpha_0-\boldsymbol{\alpha}_L\cdot \mathbf{E}_L(\mathbf{x}_t)}\mathrm{d}\mathbf{x}_t \\
    &= -\big(\nabla_{\boldsymbol{\delta}}\boldsymbol{\alpha}_L\big)^\mathtt{T}\int_{\mathbf{x}_t}\mathbf{E}_L(\mathbf{x}_t)p^{\delta}_L(\mathbf{x}_t)\mathrm{d}\mathbf{x}_t \\
    &= -\big(\nabla_{\boldsymbol{\delta}}\boldsymbol{\alpha}_L\big)^\mathtt{T}\,\overline{\mathbf{E}}_L,
\end{align*}
As for the $L\times N$ Jacobian matrix $\nabla_{\boldsymbol{\delta}}\boldsymbol{\alpha}_L$, with standard element $\displaystyle\left(\nabla_\delta\boldsymbol{\alpha}_L\right)^{j=1,\ldots,N}_{l=1,\ldots,L}=\frac{\partial \alpha_l}{\partial \delta_j}$, we can get an expression for it implicitly by differentiating $\overline{\mathbf{E}}_L$ with respect to $\boldsymbol{\delta}$, to end up with
\begingroup
\allowdisplaybreaks
\begin{align*}
    \nabla_{\boldsymbol{\delta}}\overline{\mathbf{E}}_L&= \nabla_{\boldsymbol{\delta}}\left(\int_{\mathbf{x}_t}\mathbf{E}_L(\mathbf{x}_t)p_L^{\delta}(\mathbf{x}_t)\mathrm{d}\mathbf{x}_t\right) \\
    &= \int_{\mathbf{x}_t}\mathbf{E}_L(\mathbf{x}_t)\left(-\nabla_{\boldsymbol{\delta}}\alpha_0-\nabla_{\boldsymbol{\delta}}\left(\boldsymbol{\alpha}_L\cdot \mathbf{E}_L(\mathbf{x}_t)\right) \right)^\mathtt{T} p_L^{\delta}(\mathbf{x}_t)\mathrm{d}\mathbf{x}_t \\
    &= -\int_{\mathbf{x}_t}\mathbf{E}_L(\mathbf{x}_t)\left(-\big(\nabla_{\boldsymbol{\delta}}\boldsymbol{\alpha}_L\big)^\mathtt{T}\,\overline{\mathbf{E}}_L+\big(\nabla_{\boldsymbol{\delta}}\boldsymbol{\alpha}_L\big)^\mathtt{T}\mathbf{E}_L(\mathbf{x}_t)\right)^\mathtt{T} p_L^{\delta}(\mathbf{x}_t)\mathrm{d}\mathbf{x}_t \\
    &= -\left[\int_{\mathbf{x}_t}\mathbf{E}_L(\mathbf{x}_t)\left(\mathbf{E}_L(\mathbf{x}_t)-\overline{\mathbf{E}}_L\right)^\mathtt{T} p_L^{\delta}(\mathbf{x}_t)\mathrm{d}\mathbf{x}_t\right]\nabla_{\boldsymbol{\delta}}\boldsymbol{\alpha}_L \\
    &= -\left[\int_{\mathbf{x}_t}\left(\mathbf{E}_L(\mathbf{x}_t)-\overline{\mathbf{E}}_L\right)\left(\mathbf{E}_L(\mathbf{x}_t)-\overline{\mathbf{E}}_L\right)^\mathtt{T} p_L^{\delta}(\mathbf{x}_t)\mathrm{d}\mathbf{x}_t\right]\nabla_{\boldsymbol{\delta}}\boldsymbol{\alpha}_L \\
    &\equiv -\mathcal{C}\nabla_{\boldsymbol{\delta}}\boldsymbol{\alpha}_L.
\end{align*}
\endgroup
If we now assume invertibility of the (climate) correlation matrix $\mathcal{C}$ we have,
\begin{equation*}
    \nabla_{\boldsymbol{\delta}}\boldsymbol{\alpha}_L=-\mathcal{C}^{-1}\nabla_{\boldsymbol{\delta}}\overline{\mathbf{E}}_L.
\end{equation*}
Plugging in the expressions above for $\nabla_{\boldsymbol{\delta}}\alpha_0$ and $\nabla_{\boldsymbol{\delta}}\boldsymbol{\alpha}_L$ into $\left((\boldsymbol{\delta}\cdot\nabla_{\boldsymbol{\delta}})\left(\alpha_0+ \boldsymbol{\alpha}_L\cdot \mathbf{E}_L(\mathbf{x}_t)\right) \right)^2p_L^{\delta}(\mathbf{x}_t)$ and using the definition of the Fisher information matrix and symmetricity of the correlation matrix $\mathcal{C}$ and its inverse, we then have the following series of equalities, where all gradients are calculated at the unperturbed state of $\boldsymbol{\delta}=\boldsymbol{0}$,
\begin{align*}
    \boldsymbol{\delta}\cdot I(p(\mathbf{x}_t))\boldsymbol{\delta}&=\boldsymbol{\delta}\cdot\left(\int_{\mathbf{x}_t}\left(-\nabla_{\boldsymbol{\delta}}\alpha_0-\nabla_{\boldsymbol{\delta}}\left(\boldsymbol{\alpha}_L\cdot \mathbf{E}_L(\mathbf{x}_t)\right) \right)\left(-\nabla_{\boldsymbol{\delta}}\alpha_0-\nabla_{\boldsymbol{\delta}}\left(\boldsymbol{\alpha}_L\cdot \mathbf{E}_L(\mathbf{x}_t)\right) \right)^\mathtt{T} p_L^{\delta}(\mathbf{x}_t)\mathrm{d}\mathbf{x}_t\right)\boldsymbol{\delta} \\
    &=\boldsymbol{\delta}\cdot\left(\int_{\mathbf{x}_t} \big(\nabla_{\boldsymbol{\delta}}\overline{\mathbf{E}}_L\big)^\mathtt{T}\mathcal{C}^{-1}\left(\mathbf{E}_L(\mathbf{x}_t)-\overline{\mathbf{E}}_L\right)\left(\mathbf{E}_L(\mathbf{x}_t)-\overline{\mathbf{E}}_L\right)^\mathtt{T}\mathcal{C}^{-1}\nabla_{\boldsymbol{\delta}}\overline{\mathbf{E}}_L   p_L^{\delta}(\mathbf{x}_t)\mathrm{d}\mathbf{x}_t\right)\boldsymbol{\delta} \\
    &=\boldsymbol{\delta}\cdot\left(\big(\nabla_{\boldsymbol{\delta}}\overline{\mathbf{E}}_L\big)^\mathtt{T}\mathcal{C}^{-1}\nabla_{\boldsymbol{\delta}}\overline{\mathbf{E}}_L\right)\boldsymbol{\delta},
\end{align*}
which proves \eqref{quadratic_form_approx}. The result and proof of this theorem generalise mutatis mutandis to the case of models with complex-valued processes and complex-valued statistical quantities of interest, by replacing the regular transpose with the conjugate transpose, since the climate correlation matrix (and by extension its inverse) is, in general, Hermitian. 

\bibliographystyle{ametsocV6}
\bibliography{references}

\end{document}